\documentclass{article}

\usepackage{arxiv}
\usepackage[utf8]{inputenc} 
\usepackage[T1]{fontenc}    
\usepackage{hyperref}       
\usepackage{url}            
\usepackage{booktabs}       
\usepackage{amsfonts}       
\usepackage{nicefrac}       
\usepackage{microtype}      
\usepackage{lipsum}
\usepackage{graphicx}
\usepackage{subfigure}
\usepackage{amssymb}
\usepackage{mathtools}
\usepackage{multirow}
\usepackage{appendix}
\usepackage{indentfirst}
\usepackage{color}
\usepackage{xcolor}
\usepackage{float}

\title{Bayesian Quantile Factor Models}

\author{
 Kelly C. M. Gonçalves \\
  Departamento de Métodos Estatísticos\\
  Universidade Federal do Rio de Janeiro\\
  Av. Athos da Silveira Ramos, 149, Rio de Janeiro, RJ, Brasil \\
  \texttt{kelly@dme.ufrj.br} \\
   \And
 Afonso C. B. Silva \\
  Departamento de Métodos Estatísticos\\
  Universidade Federal do Rio de Janeiro\\
  Av. Athos da Silveira Ramos, 149, Rio de Janeiro, RJ, Brasil \\
  \texttt{afonso@dme.ufrj.br} \\
}

\begin{document}
\maketitle
\begin{abstract}
Factor analysis is a  flexible technique for assessment of multivariate dependence and codependence. Besides being an exploratory tool used to reduce the dimensionality of multivariate data, it allows estimation of common factors that often have an interesting theoretical interpretation in real problems. However, in some specific cases the interest involves the effects of latent factors not only in the mean, but in the entire response distribution, represented by a quantile. This paper introduces a new class of models, named quantile factor models, which combines factor model theory with distribution-free quantile regression producing a robust statistical method. Bayesian estimation for the proposed model is performed using an efficient Markov chain Monte Carlo algorithm. The proposed model is evaluated using synthetic datasets in different settings, in order to evaluate its robustness and performance under different quantiles compared to more usual methods. The model is also applied to a financial sector dataset and a heart disease experiment.
\end{abstract}


\section{Introduction}

Technological advancement and the strengthening of computational tools have led to increasing of large amounts of data available, making dimensionality reduction an essential aspect of data analysis. Among the multivariate analysis techniques for that purpose is factorial analysis, which aims to describe the original dependence of a set of  observed correlated variables by a smaller number of unobservable latent variables called latent factors.

Factor models are structured from a linear model that relates observable variables to the latent factors plus a random error component. The dependence structure of the original variables is explained by the matrix of factor loadings. Although in standard factor analysis, a multivariate normal distribution is assumed for the errors \cite{lopes2004bayesian}, the literature has introduced various factor models, such as a general class of factor-analytic models for the analysis of multivariate (truncated) count data \cite{wedel2003factor}, binary data coming from an unobserved 
heterogeneous population \cite{cagnone2012factor}, robust analysis replacing the Gaussian factor analysis model with a multivariate Student-t distribution \cite{zhang2014robust}, censored non-normal random variables with influential observations \cite{castro2015likelihood}, and categorical variables \cite{CapdevilleGoncPer}. For all these cases, the latent factors are a representation of the original variables, obtained from the expected value, which entails two possibly restrictive features:  (i) hidden factors that may shift characteristics (moments or quantiles) of the distribution of the original variable rather than its mean are not captured; and (ii) the factor loadings are not allowed to vary across the distributional characteristics of each unit.

Interest in investigating associations of random variables in tail parts has arisen in various fields. In finance, for example, recurrent global finance crises have shown that the risky status, or Value at Risk (VaR), of one financial institution may cause a series of negative impacts on other financial institutions or the entire financial system \cite{tobias2016covar}.  In environmental science, the growing frequency of abnormal climate events has increased the importance of identifying associations of environmental factors in the extreme tail part of the distribution \cite{villarini2011frequency}. In particular, in this work, one of the applications discussed in Section \ref{sec4} shows how the association between several world market indices may be influenced when analyzing lower quantiles, for example. More specifically, the effect of the Russell 2000 index of the New York Stock Exchange on other market indices is greater when observed at the 10\% quantile than at the expected value.

In the context of regression analysis, quantile regression has been an appealing application \cite{koenker1978regression} in cases where the interest relies in the effects of covariates not only in the mean, but in the entire response distribution, represented by a quantile. Quantile regression
is advantageous compared to standard mean regression because besides providing richer information about the covariate effects, it is robust to heteroscedasticity and outliers, and accommodates the non-normal errors often encountered in practical applications.

Bayesian inference for quantile regression operates by
forming the likelihood function based on the asymmetric Laplace distribution \cite{yu2001bayesian}. In particular, \cite{kozumi2011gibbs} proposed a Gibbs sampling algorithm to sample
from the posterior distribution of the unknown quantities using a normal-exponential location-scale
mixture representation for the asymmetric Laplace distribution.

Based on these ideas, the main aim of this work is to extend the standard factor models to a flexible new class, named quantile factor models, which allows capturing both the quantile-dependent loadings and extra latent factors.
In the proposed method a linear function of the latent factors is set equal to a quantile of the
response variables, similar to the quantile regression of \cite{koenker1978regression}. Besides being an exploratory tool used to reduce the dimensionality of multivariate data based on a quantile correlation structure, the method allows estimation of common factors that often have an interesting theoretical interpretation in real problems and that may vary depending on the quantile. 

To follow the Bayesian paradigm, like \cite{yu2001bayesian}, a multivariate asymmetric Laplace distribution is assumed for the random errors of the proposed factorial model. Since the kernel of the posterior distribution does not result in a known distribution, we make use of Markov chain Monte Carlo (MCMC) methods to sample from it.

The remainder of the paper is organized as follows. Section \ref{sec-mfq} presents the proposed model. It starts by discussing some aspects of the method when applied to the simplest case, namely the univariate one assuming the latent factors to be fixed. Then, the proposed method is introduced as an extension of those ideas to the multivariate case and assuming unknown factors. Some model properties are discussed together with the inference procedure, which follows the Bayesian paradigm and makes use of Gibbs sampling for almost all the parameters, since the posterior full conditional distributions have a closed analytical form, except for some scale parameters, for which the Metropolis-Hastings algorithm is used. Section \ref{sec3} presents the analysis of two synthetic datasets. First a dataset is generated from a transformation of a Normal factor model, for which correlation seems to be larger for upper quantiles. Then, the results obtained in the quantile factor model fit for some quantiles  are compared to those obtained under Normal and Student-t \cite{zhang2014robust} factor model fits. The second illustration consists of a comparison of the results obtained with the proposed model in the median with the Normal and Student-t factor models using a synthetic dataset with outliers, in order to study the robustness of the method. In Section \ref{sec4}, 
we illustrate the approach for two real datasets. Model comparison is performed using different criteria and we concentrate on the practical interpretations using the factor loadings estimates. Finally, Section \ref{sec5} discusses our main findings. 

\section{Quantile factor model}\label{sec-mfq}

Let a random $p$-vector $y_i$, $i=1,\dots,n$, be a measurement of $k$ ($k \leq p$) latent factors, $f_i^{(\tau)} = (f_{1i}^{(\tau)}, \dots, f_{ki}^{(\tau)})$. The proposed model assumes that the associations among the observed variables in $\tau$-th quantile ($0<\tau<1$) are wholly explained by the $q$ latent variables, such that: 
\begin{equation}\label{eq1}
    \mathcal{Q}_\tau(y_i | f_i) =  \beta_\tau f_i^{(\tau)},
\end{equation}
where $\mathcal{Q}_\tau(y_i | f_i)$ is the $\tau$-th $p$-quantile of $y_i$, formally defined as $\mathcal{Q}_\tau(y_i) = \inf \{y^* : P(y_i < y^* ) \geq \tau\}$, for $y^*\in \mathbb{R}^p$, $\beta_\tau$ is the $p\times k$ $\tau$-factor loadings matrix analogous to factor loadings in standard factor models. Besides the factor loadings, the method allows factors to exhibit heterogeneous effects across different parts of the conditional distribution.  

Conditioning equation (\ref{eq1}) on the latent factors and assuming $p=1$, we get $y_i$ a scalar, so the problem described can be viewed as a standard quantile regression. To motivate the proposed model's construction, we first describe its presentation first describing this particular case. From now on, the sub and superscript $\tau$ will be omitted in order to keep the notation as simple as possible.

\subsection{A particular introductory case}\label{univ}

 The standard quantile regression is obtained conditionally on the latent factors, assuming $p=1$ and adding to the term $\beta f_i$ in equation (\ref{eq1}) an error term $\epsilon_i$  whose distribution (with density, say, $g_\tau(.)$) is restricted so that the $\tau$-th quantile is equal to zero, that is $\int_{-\infty}^{0} g_\tau(\epsilon_i) d\epsilon_i = \tau$. For this case,
\cite{koenker1978regression} defined the $\tau$-th quantile regression estimator of $\beta$ as any solution of the following quantile minimization problem:
\begin{equation*}
\min_{\beta}\sum_{i=1}^n{\phi_\tau\left(y_i-\beta f_i\right)},
\end{equation*}
where $\phi_\tau(.)$ is the loss (or check) function defined by $\phi_\tau(u) = u( \tau - I(u<0)),$ with $I(\cdot)$ denoting the indicator function. 

In this particular case, the Bayesian inference for quantile regression proceeds by using the idea that minimizing the loss function $\phi_\tau(\cdot)$ is equivalent to maximizing the likelihood function of an asymmetric Laplace ($\mathcal{AL}$) distribution \cite{yu2001bayesian}. However, instead of maximizing the likelihood, \cite{yu2001bayesian} obtained the posterior distribution of the $\tau$-th quantile regression coefficients using the $\mathcal{AL}$ distribution. Thus, to use the Bayesian inference paradigm for quantile regression, it is enough to assume that, regardless of the distribution of $y_i$, $\epsilon_i\sim \mathcal{LA}(0, \sigma, \tau)$, whose density function is given by:
\begin{equation}\label{eq:fdp-laplace-assimetrica-univariada}
    g_\tau(\epsilon_i) = \frac{\tau(1-\tau)}{\sigma}\exp \left( -\frac{\phi_\tau(\epsilon_i)}{\sigma} \right),
\end{equation}
with $\epsilon_i \in \mathbb{R}$, $0 < \tau < 1$ a skewness parameter representing the quantile of interest and $\sigma > 0$ a scale parameter. Conditional on $f_i$, we have that $y_i\mid f_i\sim \mathcal{AL}(\beta f_i, \sigma, \tau)$.

\cite{kotz2012laplace} presented a location-scale mixture representation of the $\mathcal{AL}$ distribution that allows finding analytical expressions for the conditional posterior densities of the model. Then, conditional on $f_i$ we can write the distribution of $y_i$ using the following mixture representation:
\begin{equation}\label{eq:mistura-laplace-assimetrica-univariada}
    y_i = \beta f_i + a_\tau w_i  + b_\tau \sqrt{\sigma w_i} v_i, 
\end{equation}
for $v_i \sim N(0, 1)$, $w_i \sim Exp(\sigma)$, where $Exp(\lambda)$ denotes the Exponential distribution with mean $\lambda$,
\begin{equation*}
a_\tau = \frac{1 - 2\tau}{\tau(1-\tau)} \hspace{0.4cm} \textrm{and} \hspace{0.4cm} b^2_\tau = \frac{2}{\tau(1-\tau)}. 
\end{equation*} 

Moreover, its mean and variance are, respectively, given by:
\begin{equation*}
E(y_i) = \beta f_i + \sigma a_\tau
\hspace{0.4cm} \textrm{ and } \hspace{0.4cm}
Var(y_i) =  \sigma^2 (a_\tau^2+ b_\tau).
\end{equation*}	

The quantile factor model, presented next, can be viewed as a multivariate extension ($p>1$) of model in (\ref{eq:mistura-laplace-assimetrica-univariada}), assuming also that $f_i$ is unknown.


\subsection{Proposed model}

The quantile factor model relates each $p$-vector $y_i$ to the underlying $k$-vector of random variables $f_i$ such that: for $i=1,\dots,n$
	\begin{equation}\label{eq:definicao-mfq}
	y_i = \beta f_i + \epsilon_i, \;\; \epsilon_i \sim \mathcal{AL}_p(m, \Delta),
	\end{equation}
where $f_i$ is independent with $f_i\sim N_k(0,I_k)$, for $I_k$ an identity $k$-matrix; $\mathcal{AL}_p(\mu,\Psi)$ denotes the $p-$multivariate asymmetric Laplace distribution with $\mu \in \mathbb{R}^p$ and $\Psi$ is a symmetric positive-definite $p \times p$-matrix \cite[ch. 6]{kotz2012laplace}; $\epsilon_i$ and $f_s$ are independent for all $i$ and $s$. A brief presentation of the $p-$multivariate asymmetric Laplace ($\mathcal{AL}_p$) distribution, its moments and the characteristic function are presented in Appendix A.

From the definition in equation (\ref{eq:definicao-mfq}) and the marginal properties presented in Appendix A, we get, conditional on latent factors, that for all $l,h=1,\dots,p$,
\begin{align*}
Cov(y_{li}, y_{hi} | f_i ) &=   m_lm_h + \delta_{lh}=\left(\frac{1-2\tau}{\tau(1-\tau)}\right) ^ 2 \sigma_l \sigma_h + \delta_{lh} \mbox{ and}\\
    Var(y_{li} | f_i) &= m^2_l+\delta_{ll} = \sigma_l^2 \frac{(1-2\tau + 2\tau^2)}{\tau^2(1-\tau)^2}.
\end{align*}

Analogously, the pairwise covariance and variance, marginal on latent factors, are, respectively, given by:
\begin{align}
	Cov(y_{li}, y_{hi})  =  & \displaystyle  \sum_{j=1}^{k} \beta_{lj} \beta_{hj} + m_lm_h + \delta_{lh} = \displaystyle  \sum_{j=1}^{k} \beta_{lj} \beta_{hj} + \left(\frac{1-2\tau}{\tau(1-\tau)}\right)^2 \sigma_l \sigma_h + \delta_{lh}\nonumber\\
	\mbox{and } Var(y_{li})  = & \sum_{j = 1}^{k} \beta_{lj}^2 + m^2_l + \delta_{ll}  =  \displaystyle \sum_{j = 1}^{k} \beta_{lj}^2 + \sigma_l^2 \frac{(1-2\tau + 2\tau^2)}{\tau^2(1-\tau)^2}.\label{eq2}
	\end{align}
According to (\ref{eq2}), we have that the variance is divided into a part explained by the common factors and the uniquenesses, which measure the residual variability for each of the variables once those contributed by the factors are accounted for. This feature is in agreement with the normal factor model.	

Note that while in the normal factor model, the conditional covariance is always zero, this only happens in the proposed model in (\ref{eq:definicao-mfq}) if we assume $\delta_{lh} = 0$ and $\tau = 1/2$. Thus, we particularly assume in this paper that $\delta_{lh}=0$, in order to have an approach closest to the standard theory. 

The variance decomposition, which is a fairly standard way to summarize the importance of a common factor by its percentage contribution to the variability of a given attribute, is given by:
\begin{equation}\label{DV}
DV_l = 100\frac{ \displaystyle \sum_{j = 1}^{k} \beta_{lj}^2}{ \displaystyle \sum_{j = 1}^{k} \beta_{lj}^2 + \sigma_l^2 \frac{(1-2\tau + 2\tau^2)}{\tau^2(1-\tau)^2}}\%,
\end{equation} 
for $l = 1,\dots, p.$  Since the uniqueness $\sigma^2_l$ is multiplied by a factor that depends on $\tau$, the variance decomposition in (\ref{DV}) should be carefully used, mainly when the interest lies in comparing different quantiles or the proposed model with other ones, whose penalty is different. The uniqueness is increased eightfold if the median is tracked,  when compared to the variance decomposition in the Normal factor model, and this inflation becomes larger as the quantile moves to the distribution's tails. 



Moreover, as happens in the normal factor model, the proposed model (\ref{eq:definicao-mfq}) suffers from a identifiability problem, due to the invariance of factor models under orthogonal transformations. 
That is, if one considers $\beta^*= \beta \Gamma^{-1}$ and $f_i^*= \Gamma f_i$, for any nonsingular matrix $\Gamma$, the same model defined in (\ref{eq:definicao-mfq}) is obtained. To deal with the factor model invariance, the alternative adopted here to identify the model is to constrain $\beta$ to be a block lower 
triangular matrix, assumed to be of full rank, with strictly positive diagonal elements. This form provides both identification and often useful interpretation of the factor model \cite{lopes2004bayesian}. 

With a specified $k$-factor model, Bayesian analysis using MCMC methods is straightforward. \cite{lopes2004bayesian} treated the case where uncertainty about the number of latent factors is assumed in a Normal factor model. They also discussed reversible jump MCMC methods and alternative MCMC methods based on bridge sampling.

In the identifiable model, the loadings matrix has
$pk - k(k-1)/2$ free parameters. With $p$ non-zero $\sigma^2_j$, $j=1,\dots,p$ parameters, the
resulting factor form of $\Delta$ has $p(k+1)-k(k-1)/2$ parameters, compared with the total $p(p + 1)/2$ in an unconstrained (or $p = k$) model; leading to the
constraint that $p(p + 1)/2 - p(k + 1) + k(k - 1)/2 \geq 0$, which provides at least an upper bound on $k$ \cite{lopes2004bayesian}. In this work, we assume $k$ known and use an exploratory pairwise quantile correlation plot as a preliminary method to infer the $k$-value. Moreover, the method is flexible enough to allow the use of different values for $k$, depending on the quantile of interest.

The choice of the quantile to be tracked in the following applications depends on the specific aims of the problem. In each example we arbitrarily fix a small quantile, 10\%, the median, and a large quantile, 90\%, to be monitored, with the purpose of illustrating the method for different scenarios and quantile correlation values. However, in some contexts, there is a practical rationale
behind this choice. For example, in the Value-at-Risk (VaR) estimation, used by financial
institutions and their regulators as the standard measure of market risk, the $\tau$-th
quantile is often set to 0.01 or 0.05. Or in survival analysis, in the mean residual life estimation this is especially useful when the tail behavior of the distribution is of interest. The same rationale applies to in the quantile factor model fit.

\subsubsection{Inference procedure}

The $\mathcal{AL}_p$ distribution admits a location-scale mixture representation that allows finding analytical expressions for the conditional posterior densities of the model \cite{kozumi2011gibbs}. 
Therefore, the quantile factor model in equation (\ref{eq:definicao-mfq}) can be rewritten as
the following hierarchical model:
\begin{align}\label{mod_repr_mist}
\begin{array}{rl}
	y_i | \beta, f_i,\Delta,\tau, w_i &\sim  N_p(\beta f_i + m w_i,  w_i \Delta),\\
	w_i &\sim  Exp(1),
	\end{array}
\end{align}
with $f_i \sim N_k(0, I_k).$  

Let $\Theta= (\beta,f_1,\dots,f_n,\sigma^2_1,\dots,\sigma^2_p, w_1,\dots,w_n)$
be the parameter vector. The inference procedure is performed under the Bayesian paradigm assuming the number of factors $k$ to be known, and model specification is completed after assigning a
prior distribution for $\Theta$, $p(\Theta)$. An advantage of following the Bayesian paradigm is that the inference procedure is performed in a single framework, and uncertainty about parameter estimation is naturally accounted for. We assume some
components of $\Theta$ are independent, {\it a priori}. More specifically,
\begin{align*}
p(\Theta)=\left[\prod_{i=1}^n{g(f_i)}\right]\left[\prod_{j=1}^p{p(\sigma^2_j)}\right]p(\beta),    
\end{align*}
where $g(f_i)$ is the pdf of the $q$-multivariate normal with all the components of the mean vector equal to 0, and correlation identity matrix. 
We assume further that {\it a priori}, $\beta_{jl} \sim N(0,C_0)$, when $j\neq l$, $\beta_{jj} \sim N(0,C_0)I(\beta_{jj}>0)$ and $\sigma_j^{2}\sim IG(\nu/2,\nu s^2/2)$, where $IG(a,b)$ denotes the inverse gamma distribution having mode $s^2$ with $\nu$ being the prior degrees of freedom hyperparameter.

Following Bayes’ theorem, the posterior distribution of $\Theta$, which is proportional to the product of the likelihood function and the prior distribution for $\Theta$, is given by:
    \begin{equation*}
	\begin{array}{ll}
	p(\Theta | y_1,\dots,y_n) & \propto   \displaystyle  \prod_{i=1}^{n} p(y_i | f_i, w_i)  p(f_i) p(w_i)
	\displaystyle  \prod_{j=1}^{k} p(\beta_j) \displaystyle  \prod_{j=k+1}^{p} p(\beta_j) \displaystyle  \prod_{j=1}^{p} p(\sigma^{-2}_j) \\ \vspace{0.2cm}
	& \propto \exp \left(-\frac{1}{2} \displaystyle \sum_{i=1}^{n}(y_i - \beta f_i - mw_i)'(w_i \Delta)^{-1}(y_i - \beta f_i - mw_i)  \right) \times \\ \vspace{0.2cm}
	& \displaystyle \prod_{i=1}^{n} |w_i \Delta|^{-\frac{1}{2}} \displaystyle  \prod_{i=1}^{n} \exp \left( -w_i \right) \times \exp \left(-\frac{1}{2} \displaystyle \sum_{i=1}^{n}f_i'I_kf_i  \right) \times \\ \vspace{0.2cm}
	& \displaystyle \prod_{j=1}^{p} (\sigma^{-2}_j)^{\frac{\nu}{2}-1} \textrm{exp} \left( - \frac{\nu s^2}{2} \sigma^{-2}_j \right) \times \\ \vspace{0.2cm}
	& \displaystyle \prod_{j=1}^k   \textrm{exp} \left( -\frac{1}{2} \beta_j (C_0  I_j)^{-1} {\beta_j}'  \mathbb{I}(\beta_{jj} \geq 0) \right)   \displaystyle \prod_{j=k+1}^p  \textrm{exp} \left( -\frac{1}{2} \beta_j (C_0I_k)^{-1} {\beta_j}' \right),
	\end{array}
	\end{equation*}
	
for $i = 1, \dots, n$, $j = 1, \dots, p$, and $\beta_j$ the $j$-th line of matrix $\beta$. Moreover, we have $f_i = (f_{i1}, \dots, f_{ik})$ and denote by $f_i^j$ a $j-$vector containing the first $j$ elements of $f_i$. 
	The kernel of this distribution does not result in a known distribution. We make use of MCMC methods to obtain samples from the resulting posterior distribution. In particular, we use the Gibbs sampling algorithm for all the parameters, except for $\sigma^2_1, \ldots, \sigma^2_p$, whose full conditional distributions do not have a closed form, so we make use of  the Metropolis-Hastings algorithm with a random walk proposal distribution to obtain samples from them. However, it is quite interesting that, when $\tau = 1/2$, 
	those full conditional distributions also have a closed form, so Gibbs sampling alone could be used. The full conditional posterior distributions are described in Appendix B.
	
	In the following applications, we also perform model comparison with different values of $q$ using the Akaike information criterion (AIC), Bayesian information criterion (BIC), a variant BIC* and the informational complexity criterion (ICOMP). Each of these criteria is described in more detail in Appendix C.

	\section{Illustrations with synthetic data}\label{sec3}
	
	In this section we analyze two synthetic datasets generated from standard models, to
check the proposed model's performance in different scenarios and under different quantiles. The main aim here is to compare the proposed model with standard models, not only with respect to the fit, but also the practical interpretations that may be deduced from each case. 

In the first case study we generated a synthetic dataset from a transformation of a multivariate Normal distribution, with the purpose of comparing the proposed model with several quantiles, denoted by QFM, with the Normal and Student-t factor models, denoted by NFM and TFM, respectively. The generation process reflects the tail-dependent degree of pairwise association. In the second study, a synthetic dataset is generated from a multivariate Student-t model, with the purpose of verifying the quantile factor model fit in the median in comparison to the fit of the Student-t and Normal factor models.

We considered the following hyperparameters in the prior distributions: $C_0=100$, $\nu=0.02$ and $s^2=1$, which yield vague priors. The MCMC algorithm was implemented in the \texttt{R} programming language, version 3.4.1 \cite{teamR}, in a computer with an Intel(R) Core(TM) 
i5-4590 processor with 3.30 GHz and 8 GB of RAM memory. For each sample and fitted model, we ran two parallel chains starting from different 
values, letting each chain run for 160,000 iterations, discarded the first 10,000 as burn-in, and stored every 50th iteration to avoid possible autocorrelations within the chains. We used the diagnostic tools available in the CODA package \cite{plummer2006coda} to check convergence of the chains.

A preliminary analysis of the artificial datasets generated is done using scatterplots and a Bayesian version of the quantile correlation measure, denoted by $\rho_\tau$, proposed by \cite{choi2018quantile} and described next.

\subsection{Bayesian quantile correlation}

The usual correlation coefficients, as the Pearson correlation coefficient, which is a measure of linear association, tend to fail in the measurement of tail-specific relationships. The quantile correlation coefficient measures tail dependence in this context. In particular, \cite{choi2018quantile} defined the quantile correlation for two random variables $x$ and $y$ by:
\begin{equation}\label{choishin}
    \rho_\tau = sign(\beta_{2.1}(\tau))\sqrt{\beta_{2.1}(\tau)\beta_{1.2}(\tau)},
\end{equation}
which is the geometric mean of the two $\tau$-quantile regression slopes $\beta_{2.1}(\tau)$ of $y$ on $x$ and $\beta_{1.2}(\tau)$ of $x$ on $y$ .

While the Pearson correlation coefficient measures sensitivity of the conditional
mean of a random variable with respect to a change in the other variable, the quantile correlation $\rho_\tau$ is modified to measure sensitivity of conditional quantiles rather than conditional mean by considering $\tau$-quantile regressions.

The quantile correlation coefficient satisfies the basic features of correlation coefficient, such as: being zero for independent random variables; being bounded by 1 in absolute value for a wide class of
distributions with $1$ and $-1$ indicating perfectly linear related random variables; and having commutativity and  scale-location-invariance. The larger the absolute value of $\rho_\tau$ is, the more sensitive the conditional $\tau-$quantile
of a random variable to change of the other variable will be. More details about the quantile correlation coefficient can be seen in \cite{choi2018quantile}.

In this work, we make use of a Bayesian version of the coefficient. With the posterior sample obtained from MCMC for $\beta_{2.1}(\tau)$ and $\beta_{1.2}(\tau)$, we can propagate them and obtain a sample of posterior distribution of $\rho_\tau$. In particular, we define the Bayesian quantile correlation estimator as the posterior mean of $\rho_\tau$.

\subsection{Case study 1}\label{studycase1}

In this illustration, a sample with $n=150$ observations was generated from a multivariate Normal distribution with $p=5$, zero mean vector and covariance matrix
$ \Psi = 
   \begin{pmatrix} 
1 & 0 & 0 & 0 & 0  \\
0 & 1 & 0 & 0 & 0  \\
0 & 0 & 1 & 0.95 & 0.95  \\
0 & 0 & 0.95 & 1 & 0.95  \\
0 & 0 & 0.95 & 0.95 & 1  \\
\end{pmatrix}.$

An auxiliary variable $e$ was generated from $N(0, 9)$ and the following transformations were applied to the original dataset:
\begin{align*}
    y_{i1} & = y_{i1} + e  \mathbb{I}(y_{i1} < c)  \mathbb{I}(y_{i2} < c) \\
    y_{i2} & = y_{i2} + e  \mathbb{I}(y_{i1} < c)  \mathbb{I}(y_{i2} < c), \,\, \mbox{ for } c = -0.4 \mbox{ and } i = 1, \dots, n.
\end{align*}

\begin{figure}[h!]
    \centering
    \includegraphics[scale = 0.65]{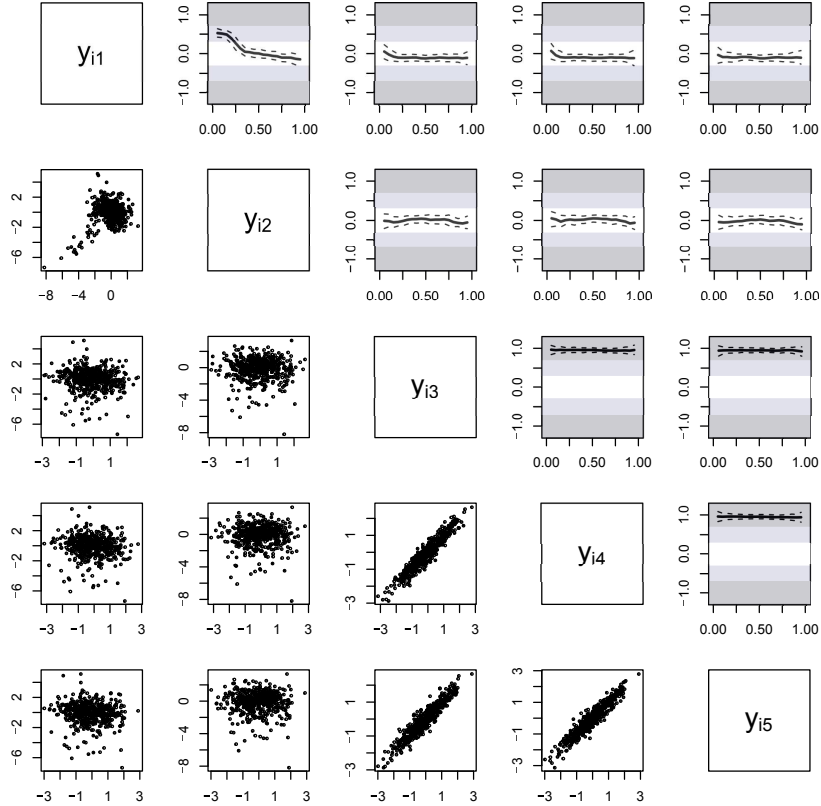}
    \caption{Lower panels: Scatterplots of dataset generated in case study 1. Upper panels: posterior mean (solid line) with respective 95\% credible interval (dashed line) of the pairwise quantile correlation varying by quantile. The gray scale highlights regions for which the correlation is weak ($|\rho_\tau| < 0.3$), moderate ($ 0.3 < |\rho_\tau| < 0.7$) and strong ($|\rho_\tau| > 0.7$). }
    \vspace{-0.3 cm}
    \label{fig:pairs-y-ilustr1}
\end{figure}

The lower panels of Figure \ref{fig:pairs-y-ilustr1} show the scatterplot for each component of the variable $y_i$, while the upper panels show the respective posterior mean (solid line) and respective 95\% credible interval (dashed line) of the pairwise quantile correlation varying by quantile. The gray scale highlights regions for which the correlation is weak ($|\rho_\tau| < 0.3$), moderate ($ 0.3 < |\rho_\tau| < 0.7$) and strong ($|\rho_\tau| > 0.7$). 

As expected from the covariance structure considered in the data generation, variables 3, 4 and 5 have strong linear and quantile correlation for all the quantiles considered. However, due to the transformation adopted, variables 1 and 2 seems to have a significant correlation only for lower quantiles and this value decreases as the quantile increases. Upper panels with the quantile correlations show that for $\tau \leq 0.3$, there is a moderate correlation between variables, while for $\tau>0.3$ it becomes weak. The other pairs seem not to be correlated  with $y_{i1}$ and $y_{i2}$, for $i = 1, \dots, n$, regardless of the quantiles considered.

\begin{figure}[h!]
    \centering
    \includegraphics[scale = 0.6]{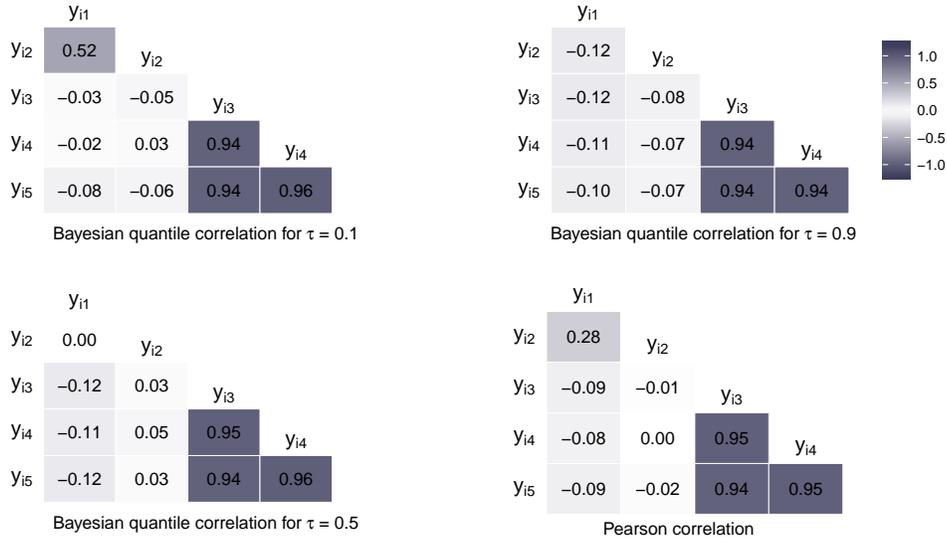}
    \vspace{-0.3 cm}
    \caption{Quantile correlation matrix estimated for $\tau=0.1, 0.5$ and $0.9$ and Pearson correlation matrix for the dataset generated in the case study 1.}
    \label{fig:matcor-datasim1}
\end{figure}

Figure \ref{fig:matcor-datasim1} presents the quantile correlation estimated  for $\tau = 0.1, 0.5, 0.9$ and the respective Pearson coefficient. While variables $y_{i3}$, $y_{i4}$ and $y_{i5}$, for $i = 1, \dots, n$, are correlated for the three quantiles considered as well as in the mean, components 1 and 2 seems to have a significant correlation just for $\tau=0.1$. Note that the Pearson correlation shows weak correlation, like an average value between the quantiles, since it is not as high as in the 0.1-th quantile and not as low as in the other quantiles.

This exploratory analysis motivates the use of the proposed quantile factor model, since the correlations are very different depending on the quantile considered. Therefore, we fit to the synthetic dataset to the proposed quantile factor model for $\tau = 0.1, 0.25, 0.5, 0.75 \textrm{ and } 0.9$, the Student-t and Normal factor model, assuming $k=1$ and $k=2$.

Table \ref{table:medidas-resumo-ilustr1} shows the different model comparison criteria computed. In a preliminary analysis the Normal factor model performs best compared to the quantile factor model with respect to the ICOMP, AIC, BIC and BIC* values. This occurs because these criteria are completely based on the likelihood function and the quantile model is not assumed to fit the data well, but to estimate relations in quantiles. Thus, we suggest the use of those comparison criteria for assessment of the quantile or the number of
factors for models in the same distribution class. For the Normal factor model, all the different criteria indicated that $k=1$ performed best among the fitted ones, while for the quantile factor model, the best results were achieved assuming $\tau=0.5$ and $k=2$. Moreover, it is quite interesting that the best value of $k$ indicated by those criteria is different depending on the quantile considered. For example, while for $\tau=0.1$ the model with $k=2$ performed better than with $k=1$, for $\tau=0.75$ this behavior changed.

On the other hand, for RPS, MAE and MSE criteria, the model that performed best was the quantile factor model assuming $\tau = 0.1$ and $k = 2$. In general, models fitted assuming $k=2$ performed better in these criteria than the models with $k=1$.

\begin{table}[h!]
\centering
\caption{Model comparison criteria for the quantile, Student-t and Normal factor models assuming $\tau = 0.1, 0.25, 0.5, 0.75 \textrm{ and } 0.9$, $k=1$ and $k=2$.\label{table:medidas-resumo-ilustr1}}
\begin{tabular}{lcccccccc}
\hline
Model & $k$ & ICOMP   & AIC     & BIC     & BIC*    & RPS           & MAE           & MSE           \\ \hline
$\textrm{QFM}_{\tau = 0.10}$   & 1   & 1690.42 & 1697.42 & 1727.53 & 1727.25 & 0.57          & 0.73          & 2.68          \\ \hline
$\textrm{QFM}_{\tau = 0.25}$   & 1   & 1576.49 & 1583.63 & 1613.74 & 1613.46 & 0.49          & 0.64          & 1.96          \\ \hline
$\textrm{QFM}_{\tau = 0.50}$   & 1   & 1442.58 & 1450.18 & 1480.28 & 1480.00 & 0.38          & 0.51          & 0.97          \\ \hline
$\textrm{QFM}_{\tau = 0.75}$   & 1   & 1530.12 & 1537.91 & 1568.02 & 1567.74 & 0.43          & 0.57          & 1.04          \\ \hline
$\textrm{QFM}_{\tau = 0.90}$   & 1   & 1635.20 & 1642.99 & 1673.09 & 1672.81 & 0.49          & 0.64          & 1.34          \\ \hline
TFM     & 1   & 391.34   & 383.85 & 353.77   & 354.05   & 0.36          & 0.46          & 0.75          \\ \hline
NFM     & 1   & 20.72   & 30.21   & 60.32   & 60.04   & 0.34          & 0.45          & 0.61          \\ \hline
$\textrm{QFM}_{\tau = 0.10}$   & 2   & 1563.63 & 1582.34 & 1624.49 & 1624.03 & 0.29 & 0.33 & 0.28 \\ \hline
$\textrm{QFM}_{\tau = 0.25}$   & 2   & 1639.70 & 1644.99 & 1687.14 & 1686.68 & 0.32          & 0.37          & 0.93          \\ \hline
$\textrm{QFM}_{\tau = 0.50}$   & 2   & 1373.74 & 1384.87 & 1427.01 & 1426.56 & 0.32          & 0.42          & 0.59          \\ \hline
$\textrm{QFM}_{\tau = 0.75}$   & 2   & 1560.27 & 1571.71 & 1613.86 & 1613.40 & 0.34          & 0.41          & 0.70          \\ \hline
$\textrm{QFM}_{\tau = 0.90}$   & 2   & 1631.60 & 1646.15 & 1688.30 & 1687.84 & 0.34          & 0.37          & 0.41          \\ \hline
TFM     & 2   & 86.97   & 96.21   & 138.38   & 137.90   & 0.35          & 0.44          & 0.67 \\ \hline
NFM     & 2   & 23.65   & 38.36   & 80.51   & 80.05   & 0.31          & 0.36          & 0.34          \\ \hline
\end{tabular}
\end{table}

\begin{figure}[h!]
    \centering
    \subfigure[$k=1$]{\includegraphics[scale = 0.22]{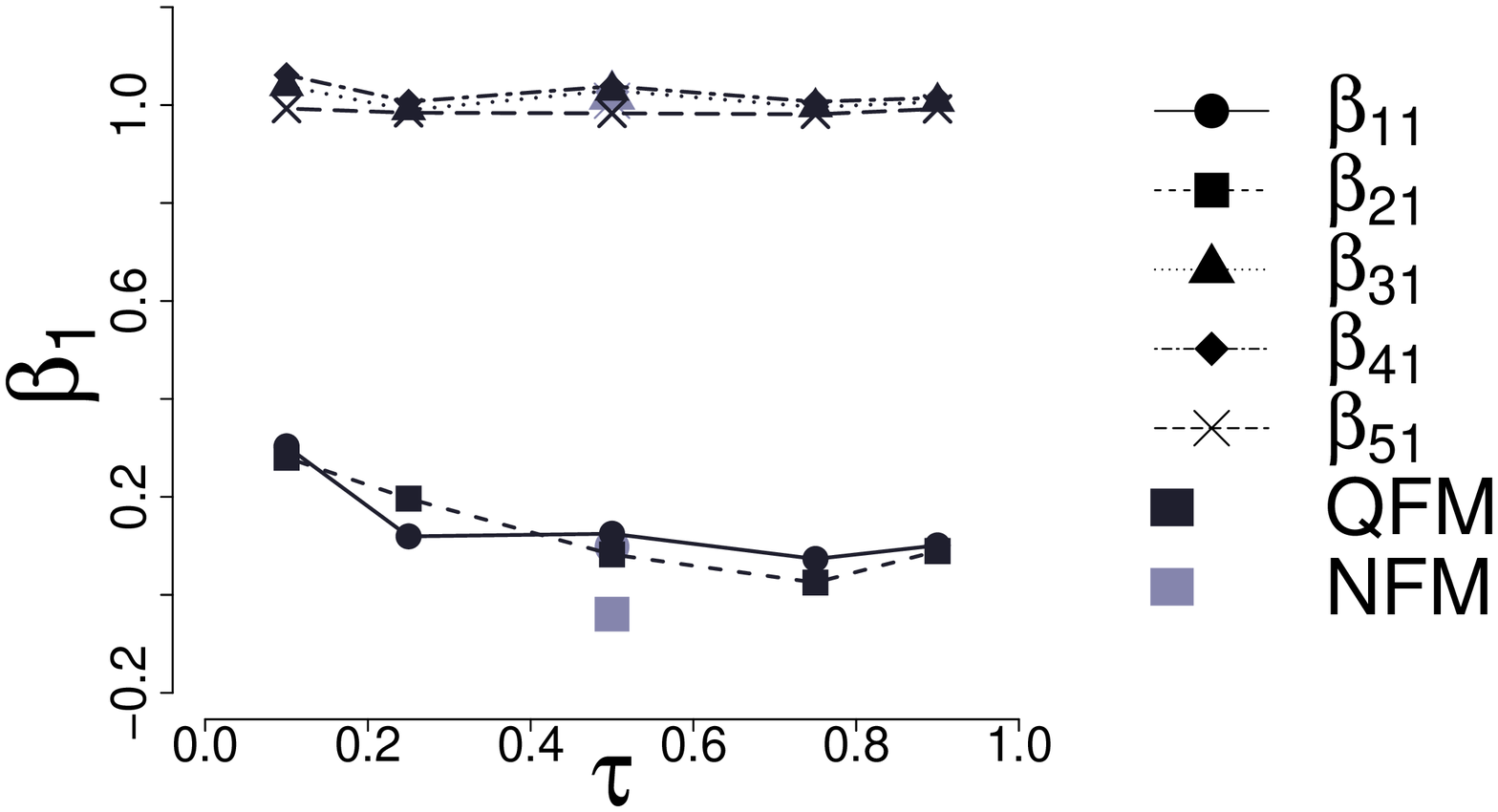}}
    \subfigure[$k=2$]{\includegraphics[scale = 0.27]{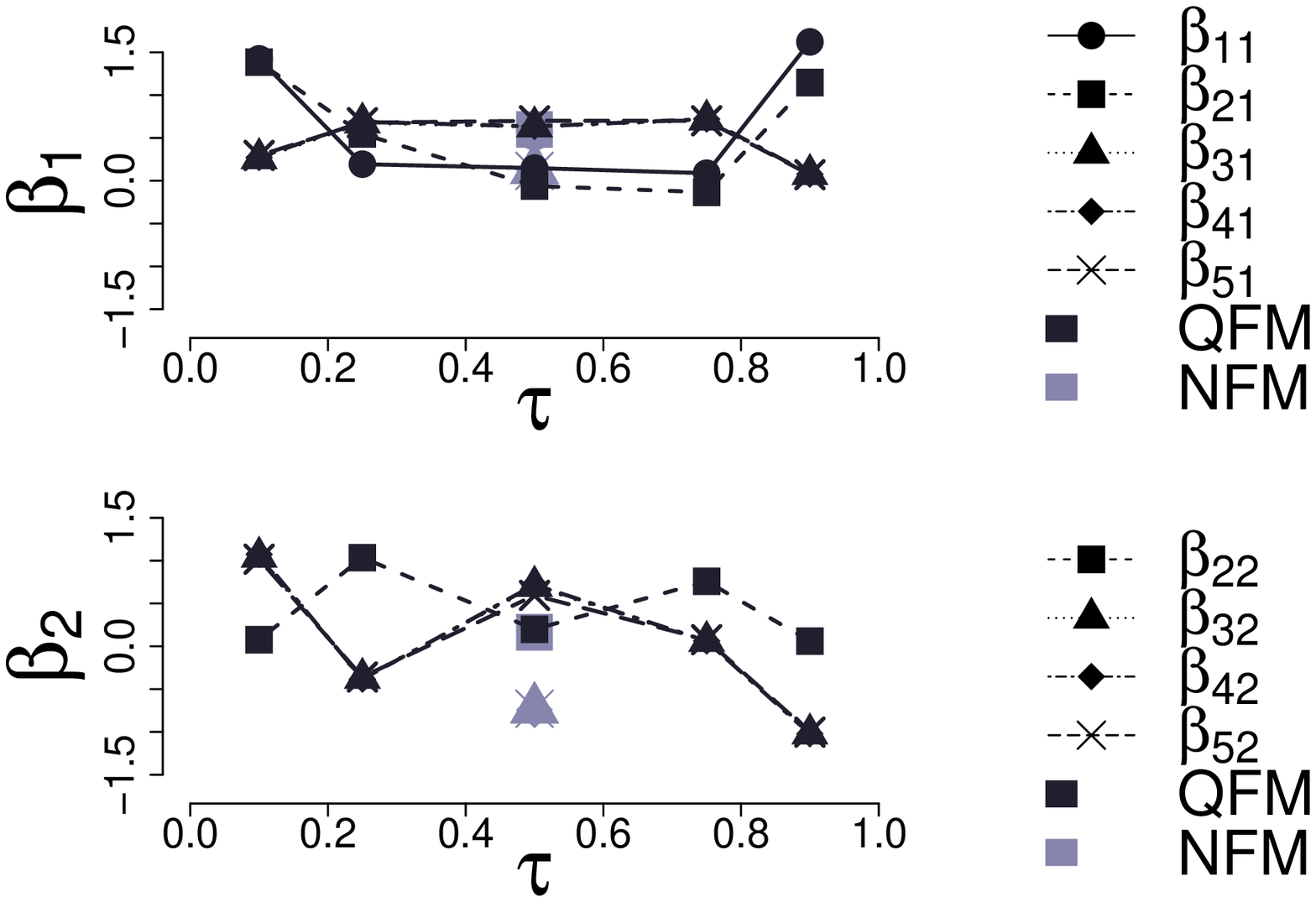}}
    \caption{Posterior mean of the factor loadings matrix obtained under the quantile factor model fit assuming $\tau = 0.1, 0.25, 0.5, 0.75 \textrm{ and } 0.90$, $k=1$ and $k=2$ and the Normal factor model.}
    \label{fig:mat-beta-datasim1-k1}
\end{figure}

Figure \ref{fig:mat-beta-datasim1-k1} displays the posterior mean of the factor loadings for the quantile factor model with varying $\tau$, for $k=1$ (a) and $k=2$ (b). The factor loadings estimated in the Normal factor model fit for $k=1$ and $k=2$ are presented in gray in the figure just for comparison.

When $k=1$ is fixed, independently of the value of $\tau$, the factor loadings estimated are higher for variables $y_{i3}, y_{i4} \textrm{ and } y_{i5}$, showing that these variables are more correlated and better explained by the latent factor than variables $y_{i1} \textrm{ and } y_{i2}$. Even so, the highest factor loadings estimated for these variables happen with $\tau=0.1$, which is exactly the one with the highest value of quantile correlation, as was shown in Figure \ref{fig:pairs-y-ilustr1}.  However, Figure \ref{fig:mat-beta-datasim1-k1} (b) shows that the factor loadings matrix is more influenced by the quantile considered when $k=2$ is assumed than for $k=1$. When $\tau = 0.1$, the first factor is  clearly related to variables $y_{i1} \textrm{ and } y_{i2}$, while the second factor is related to the other variables. Thus, the quantile factor model assuming $\tau = 0.1$ captured the correlation structure, while in the Normal factor model this structure is less evident.

Therefore, the quantile factor model allows more flexibility to summarize the association between the components for any quantile considered, either through the factor loadings, the latent factor dimension, or their interpretation. 

\begin{figure}[h!]
\centering
\subfigure[$k=1$]{\includegraphics[scale = 0.25]{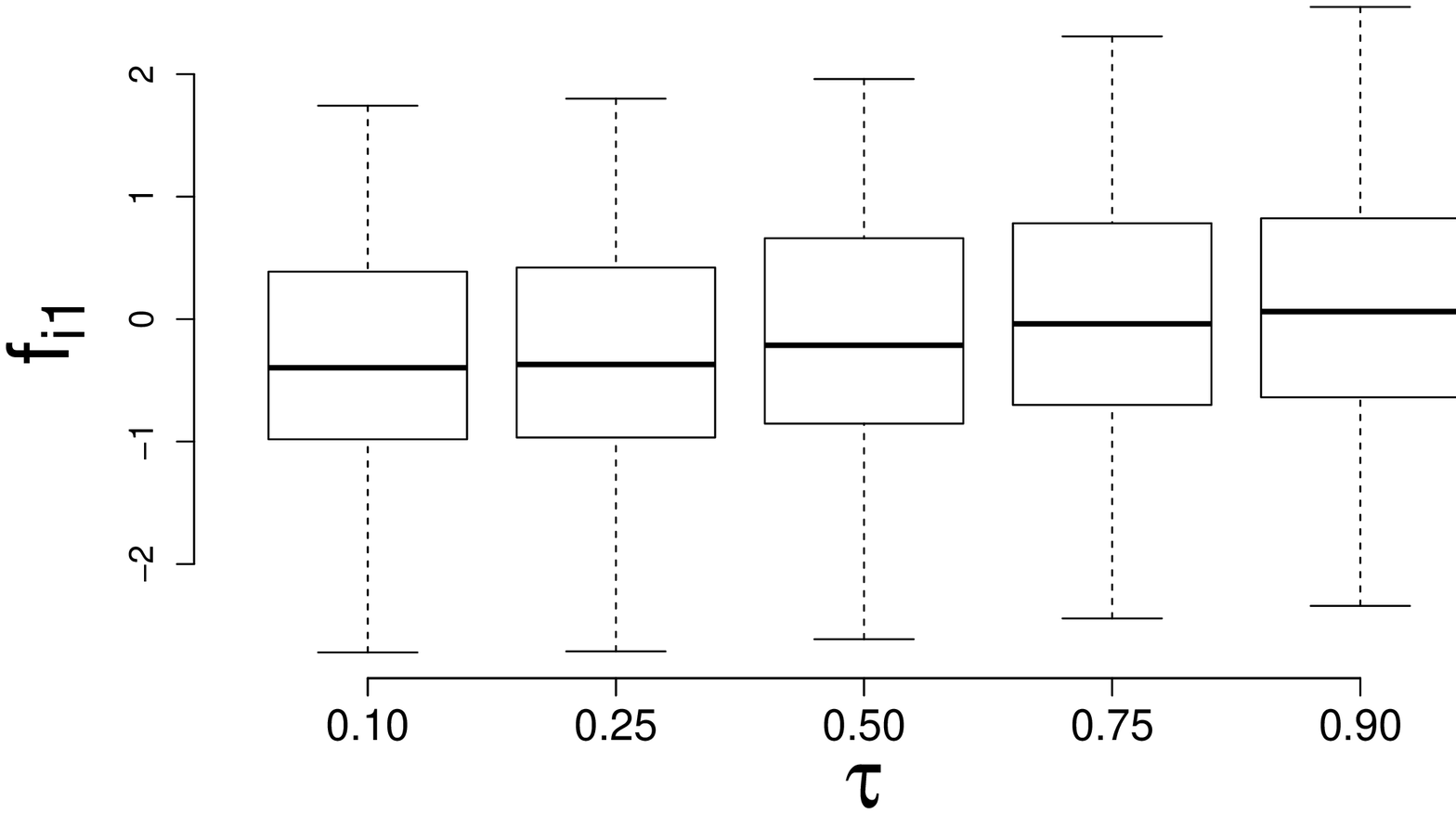}}\\
\subfigure[$k=2$]{\includegraphics[scale = 0.3]{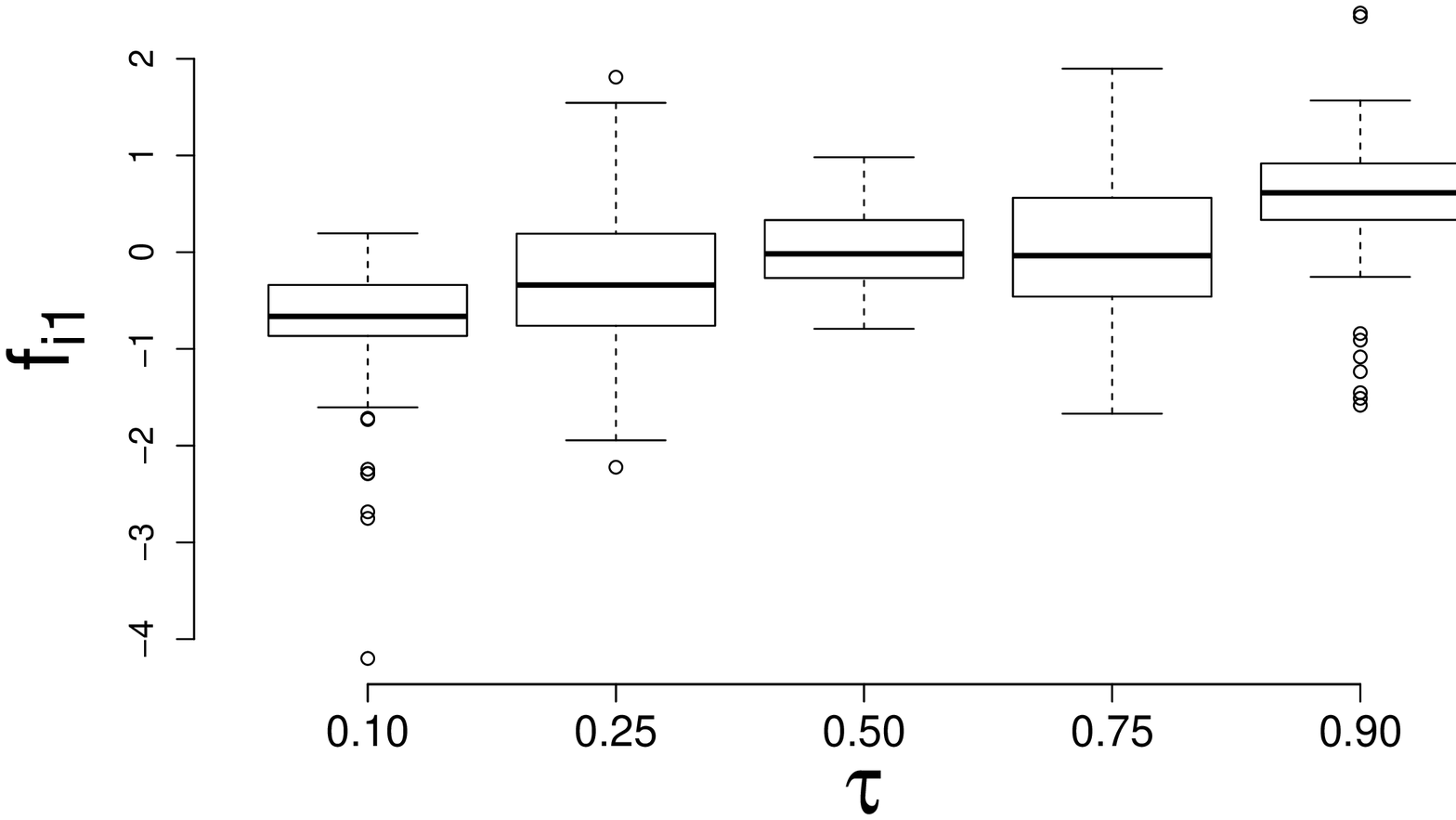}
\includegraphics[scale = 0.25]{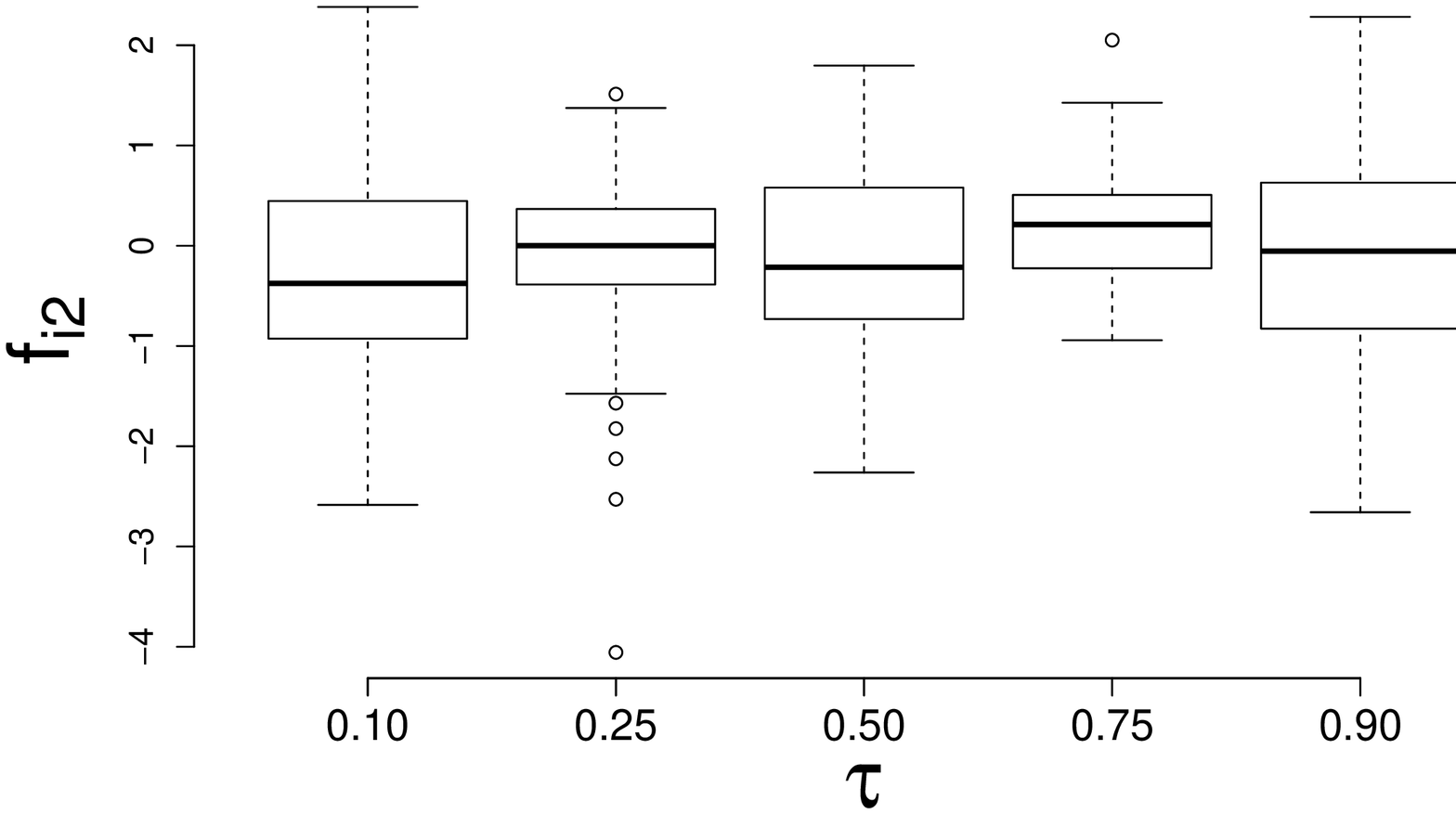}}
\caption{Boxplots of the posterior mean for the latent factors obtained in the quantile factor model fit for $\tau = 0.1, 0.25, 0.5, 0,75 \textrm{ and } 0.90$, for $k=1$ and $k = 2$.\label{fig:boxplot}}
\end{figure}

Figure \ref{fig:boxplot} displays the boxplots of the posterior mean of the latent factor obtained from the quantile factor model fit for the quantiles considered, for $k=1$ and $k=2$. In general, when $k=1$ is assumed, as the quantile increases, the factor mean increases with similar variability. However, when $k=2$, this increase is just present in the first factor, while the second factor remains at a similar level but with different variability depending on the quantile considered.  

Finally, Table \ref{dv-ilustr1} reports the posterior mean of the variance decomposition defined in equation (\ref{DV}) obtained in the quantile factor model fit assuming particularly $k=1$ and $2$ and $\tau=0.1, 0.5$ and $0.9$. We also include the variance decomposition in the Normal and Student-t factor model fits. The values are reported for each variable and separated by factor, when appropriate.

Variables 3, 4 and 5 are in general well explained for all the models considered. However, the quantile factor model with $k=2$ and $\tau = 0.1$ is the only one for which the latent factors can explain variables 1 and 2 well. The results obtained assuming $\tau=0.9$ should also be highlighted for the first variable. However, since the the variance decomposition presented in equation (\ref{DV}) contains a portion that depends on $\tau$ in the uniqueness, it must be carefully evaluated. In particular, for $\tau=0.1$ and 0.9, the uniqueness $\sigma^2_l$ is multiplied by 101 in both cases, while in the median it is just multiplied by 8. Thus, although this inflation in the uniqueness part is lower for $\tau=0.5$ than for $\tau=0.1$, the quantile factor for $\tau=0.1$ still obtains better results.

Just for comparison purposes, we include a modified version of the decomposition variance (\ref{DV}) that may be useful for the cases when comparing results with different quantiles is the interest. The modified variance decomposition for variable $l$ is defined in a similar way to the Normal factor model, as: 
\begin{equation}\label{DVmod}
DV^{\textrm{\textit{mod}}}_l = 100\frac{ \displaystyle \sum_{j = 1}^{k} \beta_{lj}^2}{ \displaystyle \sum_{j = 1}^{k} \beta_{lj}^2 + \sigma_l^2 }\%.
\end{equation}

\begin{table}[h!]
\centering
\caption{Posterior mean of the modified variance decomposition and variance decomposition obtained, respectively, in the quantile factor model fit with $\tau=0.1$, 0.5 and 0.9, and Student-t and Normal factor model fits, assuming $k=1$ and $2$.}\label{dv-ilustr1}
\vspace{0.2 cm}
\begin{tabular}{c|c|ccc|c|ccc} 
\hline
\multicolumn{1}{c}{}           & \multicolumn{4}{c|}{Variance decomposition} & \multicolumn{4}{c}{Modified variance decomposition}  \\ 
\cline{2-9}
\multicolumn{1}{c}{}           & k = 1    & \multicolumn{3}{c|}{k = 2}       & k = 1    & \multicolumn{3}{c}{k = 2}                 \\ 
\cline{2-9}
\multicolumn{1}{c}{}           & Fac 1 & Fac 1 & Fac 2 & Total      & Fac 1 & Fac 1 & Fac 2 & Total               \\ 
\hline
\multicolumn{1}{l}{}           & \multicolumn{8}{c}{$\textrm{QFM}_{\tau = 0.1}$ }                                                   \\ 
\hline
$y_{i1}$                       & 1.3      & 60.5     & 0.0      & 60.5       & 22.6     & 97.1     & 0.0      & 97.1                \\
$y_{i2}$                       & 1.5      & 64.0     & 0.2      & 64.2       & 25.6     & 97.3     & 0.3      & 97.6                \\
$y_{i3}$                       & 77.8     & 5.4      & 85.8     & 91.2       & 98.7     & 5.9      & 93.7     & 99.6                \\
$y_{i4}$                       & 78.0     & 5.5      & 86.0     & 91.5       & 98.8     & 6.0      & 93.6     & 99.6                \\
$y_{i5}$                       & 75.9     & 7.0      & 83.1     & 90.1       & 98.6     & 7.7      & 91.8     & 99.5                \\ 
\hline
\multicolumn{1}{l}{}           & \multicolumn{8}{c}{$\textrm{QFM}_{\tau = 0.5}$ }                                                   \\ 
\hline
\multicolumn{1}{l|}{$y_{i1}$ } & 0.8      & 1.1      & 0.0      & 1.1        & 6.2      & 8.3      & 0.0      & 8.3                 \\
\multicolumn{1}{l|}{$y_{i2}$ } & 0.5      & 0.3      & 2.9      & 3.2        & 4.0      & 1.8      & 19.1     & 20.9                \\
\multicolumn{1}{l|}{$y_{i3}$ } & 94.5     & 41.8     & 51.8     & 93.6       & 99.3     & 44.3     & 54.9     & 99.2                \\
$y_{i4}$                       & 93.1     & 40.6     & 52.1     & 92.7       & 99.1     & 43.4     & 55.7     & 99.0                \\
$y_{i5}$                       & 91.8     & 55.0     & 39.7     & 94.7       & 98.9     & 57.7     & 41.6     & 99.3                \\ 
\hline
\multicolumn{1}{c}{}           & \multicolumn{8}{c}{$\textrm{QFM}_{\tau = 0.9}$ }                                                   \\ 
\hline
$y_{i1}$                       & 0.1      & 60.5     & 0.0      & 60.5       & 2.2      & 97.2     & 0.0      & 97.2                \\
$y_{i2}$                       & 0.1      & 32.6     & 0.1      & 32.7       & 2.2      & 91.3     & 0.3      & 91.5                \\
$y_{i3}$                       & 78.5     & 0.5      & 88.9     & 89.4       & 98.8     & 0.5      & 98.9     & 99.5                \\
$y_{i4}$                       & 79.4     & 0.5      & 88.9     & 89.4       & 98.8     & 0.5      & 98.9     & 99.5                \\
$y_{i5}$                       & 77.2     & 0.4      & 88.3     & 88.7       & 98.7     & 0.5      & 98.9     & 99.4                \\ 
\hline
\multicolumn{9}{c}{Variance decomposition}                                                                                          \\ 
\hline
\multicolumn{1}{c}{}           & \multicolumn{4}{c|}{NFM}                    & \multicolumn{4}{c}{TFM}                              \\ 
\hline
$y_{i1}$                       & 0.6      & 20.0     & 0.0      & 20.0       & 1.5      & 2.0      & 0.0      & 2.0                 \\
$y_{i2}$                       & 0.1      & 28.7     & 2.1      & 30.8       & 0.0      & 0.0      & 3.6      & 3.6                 \\
$y_{i3}$                       & 94.7     & 2.6      & 87.9     & 90.6       & 92.5     & 75.4     & 16.1     & 91.5                \\
$y_{i4}$                       & 94.2     & 2.3      & 87.3     & 89.6       & 94.8     & 78.5     & 15.1     & 93.6                \\
$y_{i5}$                       & 94.0     & 2.2      & 87.2     & 89.3       & 92.3     & 77.6     & 13.3     & 90.9                \\
\hline
\end{tabular}
\end{table}

\clearpage

\subsection{Case study 2}

The main aim of this illustration is to evaluate the robustness of the quantile factor model for the median in the presence of outliers, compared to the Normal and Student-t factor models. With this purpose, a sample was generated from a particular multivariate Student-t model with $p=6$ variables, that is, $y \sim t_6(\mu_0, \Sigma_0, \nu)$, with $\nu = 2.5$ degrees of freedom and a shape matrix
$\Sigma_0 = 
   \begin{pmatrix} 
1 & 0.95 & 0 & 0 & 0 & 0 \\
0.95 & 1 & 0 & 0 & 0  & 0 \\
0 & 0 & 1 & 0.95 & 0 & 0 \\
0 & 0 & 0.95 & 1 & 0 & 0 \\
0 & 0 & 0 & 0 & 1 & 0.95 \\
0 & 0 & 0 & 0 & 0.95 & 1
\end{pmatrix}.$

\begin{figure}[h!]
    \centering
    \includegraphics[scale = 0.68]{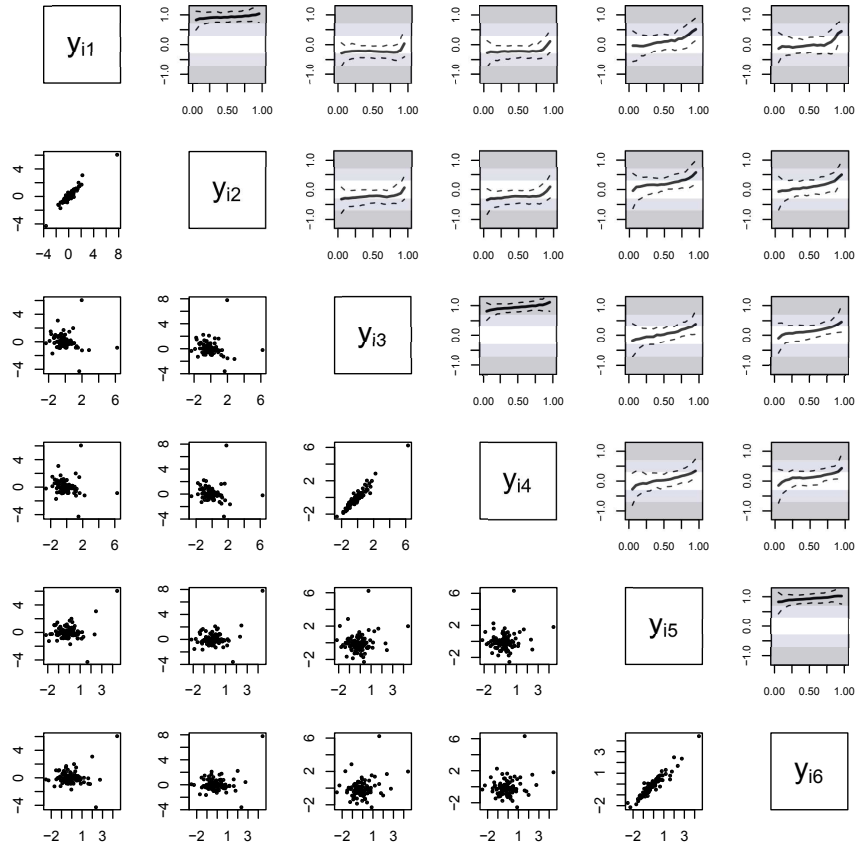}
    \caption{Lower panels: Scatterplots of dataset generated in case study 2 . Upper panels: posterior mean (solid line) with respective 95\% credible interval (dashed line) of the pairwise quantile correlation varying by quantile. The gray scale highlights regions for which the correlation is weak ($|\rho_\tau| < 0.3$), moderate ($ 0.3 < |\rho_\tau| < 0.7$) and strong ($|\rho_\tau| > 0.7$). }
    \label{fig:pairs-y-datasim3t}
\end{figure}

In a preliminary exploratory analysis, the scatterplots and quantile correlations estimated for several values of $\tau$, presented in Figure \ref{fig:pairs-y-datasim3t}, show some evidence that three latent factors are necessary to explain the variability, since there are three independent pairs of variables are be correlated. Note that in general the quantile correlations do not vary among the quantiles, except for higher quantiles, which tend to increase the quantile correlations, probably due to the presence of outliers.

Thus, to the generate dataset we fit those three factor models assuming $k = 1, 2 \textrm{ and } 3$ and compared their performance in a similar way to the case study 1 in Subsection \ref{studycase1}.

Table \ref{table-medidas-resumo-datasim3t} reports the results of the model comparison criteria considered. The quantile factor model proposed is the only one in which the criteria point $k=3$ is the best. In the Normal factor model fit, for example, the AIC indicates $k=1$, while in the Student-t factor model fit it suggests $k=2$. For the number of latent factors fixed, note that RPS, MAE and MSE in general indicate either the quantile factor model for the median or the Student-t model performs best.

\begin{table}[h!]
\centering
\caption{Model comparison criteria for the quantile factor model with $\tau=0.5$, the Normal and Student-t factor models assuming $k=1, 2$ and $3$.}\label{table-medidas-resumo-datasim3t}
\vspace{0.2 cm}
\begin{tabular}{lcccccccc}
\hline
Model                      & k & ICOMP   & AIC     & BIC     & BIC*    & RPS  & MAE  & MSE  \\ \hline
$\textrm{QFM}_{\tau = 0.5}$ & 1 & 1023.44 & 1042.73 & 1073.99 & 1073.44 & 0.35 & 0.47 & 0.76 \\ \hline
NFM                          & 1 & 21.29   & 38.85   & 70.11   & 69.56   & 0.39 & 0.48 & 0.71 \\ \hline
TFM                          & 1 & 714.77  & 732.88  & 764.15  & 763.57  & 0.37 & 0.46 & 0.74 \\ \hline
$\textrm{QFM}_{\tau = 0.5}$ & 2 & 936.05  & 958.66  & 1002.95 & 1002.05 & 0.23 & 0.29 & 0.38 \\ \hline
NFM                          & 2 & 27.53   & 46.28   & 90.57   & 89.66   & 0.26 & 0.29 & 0.35 \\ \hline
TFM                          & 2 & 264.74  & 283.80  & 328.15  & 327.18  & 0.23 & 0.28 & 0.29 \\ \hline
$\textrm{QFM}_{\tau = 0.5}$ & 3 & 866.05  & 907.77  & 962.48  & 961.22  & 0.12 & 0.11 & 0.02 \\ \hline
NFM                          & 3 & 11.98   & 53.94   & 108.65  & 107.38  & 0.13 & 0.11 & 0.02 \\ \hline
TFM                          & 3 & 694.56  & 736.47  & 791.18  & 789.91  & 0.11 & 0.10 & 0.02 \\ \hline
\end{tabular}
\end{table}

The posterior mean of the factor loadings for the three models considered are presented in Table \ref{matriz-beta-datasim3t}. When models assuming $k = 1$ are fitted, the latent factor only captures the correlation structure between variables 5 and 6. In this case, the estimates obtained in the Normal fit are significantly higher than those obtained in the quantile factor model in the median and Student-t model fit, which in turn have similar results. Assuming $k=2$ is fixed, the quantile and Normal factor models capture the correlation between variables 1 and 2 in the first latent factor and between variables 5 and 6 in the second latent factor. However, the second latent factor in the Student-t factor model is associated with variables 3 and 4. 

Finally, when $k=3$ is fixed, the three pairs of correlated variables are correctly identified in both the quantile and Student-t factor models, while the factor loadings estimated in the Normal model fit do not provide a clear distinction between the three pairs of variables, allowing $y_{i3}$, $y_{i4}$, $y_{i5}$ and $y_{i6}$ to be a combination of two latent factors.

These results suggest that since the quantile factor model with $\tau=0.5$ presented similar performance to the Student-t model, it can be viewed as an efficient alternative for studying datasets with outliers, while the Normal model seems to be more inefficient otherwise. The proposed model could be even more advantageous in this case, as it does not require estimation of the degrees of freedom, which is usually a hard parameter to estimate \cite{zhang2014robust}.

\begin{table}[h!]
\begin{center}
\caption{Posterior mean of the factor loadings matrix obtained in the quantile factor model fit assuming $\tau =0.5$, and the Normal and Student-t factor models, for $k=1, 2$ and $3$.\label{matriz-beta-datasim3t}}
\vspace{0.2 cm}
\begin{tabular}{c|c|c|cc|cc|cc}
\hline
\multicolumn{3}{c|}{$k=1$}                 & \multicolumn{6}{c}{$k=2$}                                                                             \\ \hline
$\textrm{QFM}_{\tau = 0.5}$ & NFM   & TFM   & \multicolumn{2}{c|}{$\textrm{QFM}_{\tau = 0.5}$} & \multicolumn{2}{c|}{NFM} & \multicolumn{2}{c}{TFM}\\ \hline
0.06                        & 0.25 & 0.04 & 0.94                    & 0.00                   & 1.03        & 0.00      & 0.53       & 0.00 \\
0.07                        & 0.37 & 0.03 & 1.01                    & 0.05                   & 1.12        & 0.11      & 0.59       & 0.02 \\
-0.03                       & 0.20 & 0.07 & -0.30                   & 0.04                   & -0.16       & 0.25      & 0.18       & 0.61 \\
0.01                        & 0.19 & 0.08 & -0.28                   & 0.05                   & -0.14       & 0.24      & 0.18       & 0.61 \\
0.84                        & 0.90 & 0.60 & 0.20                    & 0.87                   & 0.25        & 0.86      & 0.04       & 0.05 \\
0.89                        & 0.97 & 0.62 & 0.08                    & 0.88                   & 0.17        & 0.96      & 0.03       & 0.11 \\ \hline
\multicolumn{9}{c}{$k=3$} \\ \hline
\multicolumn{3}{c|}{$\textrm{QFM}_{\tau = 0.5}$} & \multicolumn{3}{c|}{NFM} & \multicolumn{3}{c}{TFM}\\ \hline
 \multicolumn{1}{c}{1.00}          & \multicolumn{1}{c}{0.00}           & 0.00          & 1.03   & \multicolumn{1}{c}{0.00}   & \multicolumn{1}{c|}{0.00}  & \multicolumn{1}{c}{0.54}   & \multicolumn{1}{c}{0.00}   & 0.00  \\
 \multicolumn{1}{c}{1.06}            & \multicolumn{1}{c}{0.08}           & 0.00          & 1.13   & \multicolumn{1}{c}{0.03}   & \multicolumn{1}{c|}{0.00} & \multicolumn{1}{c}{0.60}  & \multicolumn{1}{c}{0.07}   & 0.00  \\
 \multicolumn{1}{c}{-0.17}           & \multicolumn{1}{c}{0.14}           & 0.74          & -0.14  & \multicolumn{1}{c}{0.34}   & \multicolumn{1}{c|}{0.42}   & \multicolumn{1}{c}{0.18}   & \multicolumn{1}{c}{0.02}    & 0.61  \\
 \multicolumn{1}{c}{-0.16}           & \multicolumn{1}{c}{0.13}           & 0.76          & -0.12  & \multicolumn{1}{c}{0.35}   & \multicolumn{1}{c|}{0.41}  & \multicolumn{1}{c}{0.18}   & \multicolumn{1}{c}{0.01}   & 0.61  \\
\multicolumn{1}{c}{0.25}            & \multicolumn{1}{c}{0.67}           & 0.02          & 0.30   & \multicolumn{1}{c}{-0.28}  & \multicolumn{1}{c|}{0.35}  & \multicolumn{1}{c}{0.02}   & \multicolumn{1}{c}{0.62}   & 0.02  \\
 \multicolumn{1}{c}{0.15}            & \multicolumn{1}{c}{0.70}           & 0.05          & 0.23   & \multicolumn{1}{c}{-0.28}  & \multicolumn{1}{c|}{0.41}  & \multicolumn{1}{c}{0.08}   & \multicolumn{1}{c}{0.66}   & 0.07 \\ \hline
\end{tabular}
\end{center}
\end{table}

Finally, from the variance decomposition, presented in Table \ref{dv-datasim3t}, it is possible to see that the quantile and Student-t factor model fits for $k=3$ show that all the variables are equally well explained. In the Normal factor model just variables 1 and 2 are well explained by only one latent factor, while variables 3 to 6 are partially explained by the three latent factors, reflecting the estimated factor loadings. 

\begin{table}[h!]
\centering
\caption{Posterior mean of the variance decomposition obtained in the quantile factor model fit with $\tau=0.5$, and Normal and Student-t factor model fits, assuming $k=1$, 2 and 3.\label{dv-datasim3t}}
\vspace{0.2 cm}
\begin{tabular}{ccccccccc}
\hline
                           & \multicolumn{1}{c|}{$k=1$}   & \multicolumn{3}{c|}{$k=2$}                     & \multicolumn{4}{c}{$k=3$}           \\ \cline{2-9} 
                           & \multicolumn{1}{c|}{Fac 1} & Fac 1 & Fac 2 & \multicolumn{1}{c|}{Total} & Fac 1 & Fac 2 & Fac 3 & Total \\ \cline{2-9} 
\multicolumn{9}{c}{Variance decomposition - $\textrm{QFM}_{\tau = 0.5}$}                                                                      \\ \hline
\multicolumn{1}{c|}{$y_{i1}$} & \multicolumn{1}{c|}{0.5}     & 90.9    & 0.0     & \multicolumn{1}{c|}{90.9}  & 93.9    & 0.0     & 0.0     & 93.9  \\
\multicolumn{1}{c|}{$y_{i2}$} & \multicolumn{1}{c|}{0.6}     & 92.5    & 0.3     & \multicolumn{1}{c|}{92.7}  & 93.2    & 0.5     & 0.0     & 93.7  \\
\multicolumn{1}{c|}{$y_{i3}$} & \multicolumn{1}{c|}{0.1}     & 10.6    & 0.2     & \multicolumn{1}{c|}{10.8}  & 4.6     & 2.9     & 82.1    & 89.5  \\
\multicolumn{1}{c|}{$y_{i4}$} & \multicolumn{1}{c|}{0.1}     & 9.2     & 0.3     & \multicolumn{1}{c|}{9.5}   & 3.5     & 2.3     & 85.5    & 91.3  \\
\multicolumn{1}{c|}{$y_{i5}$} & \multicolumn{1}{c|}{90.6}    & 4.8     & 89.4    & \multicolumn{1}{c|}{94.2}  & 10.7    & 80.5    & 0.1     & 91.3  \\
\multicolumn{1}{c|}{$y_{i6}$} & \multicolumn{1}{c|}{88.3}    & 0.8     & 88.5    & \multicolumn{1}{c|}{89.2}  & 3.6     & 82.9    & 0.5     & 87.0  \\ \hline
\multicolumn{9}{c}{Variance decomposition - NFM}                                                                                              \\ \hline
\multicolumn{1}{c|}{$y_{i1}$} & \multicolumn{1}{c|}{5.6}     & 94.3    & 0.0     & \multicolumn{1}{c|}{94.3}  & 93.7    & 0.0     & 0.0     & 93.7  \\
\multicolumn{1}{c|}{$y_{i2}$} & \multicolumn{1}{c|}{10.4}    & 93.7    & 0.9     & \multicolumn{1}{c|}{94.6}  & 94.3    & 0.1     & 0.0     & 94.4  \\
\multicolumn{1}{c|}{$y_{i3}$} & \multicolumn{1}{c|}{3.5}     & 2.2     & 5.5     & \multicolumn{1}{c|}{7.7}   & 5.0     & 30.5    & 47.1    & 82.7  \\
\multicolumn{1}{c|}{$y_{i4}$} & \multicolumn{1}{c|}{3.2}     & 1.7     & 5.1     & \multicolumn{1}{c|}{6.8}   & 3.9     & 32.1    & 46.0    & 82.1  \\
\multicolumn{1}{c|}{$y_{i5}$} & \multicolumn{1}{c|}{92.5}    & 7.3     & 85.1    & \multicolumn{1}{c|}{92.3}  & 25.5    & 22.5    & 34.4    & 82.3  \\
\multicolumn{1}{c|}{$y_{i6}$} & \multicolumn{1}{c|}{92.4}    & 2.9     & 90.6    & \multicolumn{1}{c|}{93.6}  & 13.8    & 21.8    & 45.9    & 81.5  \\ \hline
\multicolumn{9}{c}{Variance decomposition - TFM}                                                                                              \\ \hline
\multicolumn{1}{c|}{$y_{i1}$} & \multicolumn{1}{c|}{0.2}     & 80.4    & 0.0     & \multicolumn{1}{c|}{80.4}  & 82.3    & 0.0     & 0.0     & 82.3  \\
\multicolumn{1}{c|}{$y_{i2}$} & \multicolumn{1}{c|}{0.1}     & 85.5    & 0.1     & \multicolumn{1}{c|}{85.7}  & 86.9    & 1.1     & 0.0     & 88.1  \\
\multicolumn{1}{c|}{$y_{i3}$} & \multicolumn{1}{c|}{0.4}     & 7.1     & 79.3    & \multicolumn{1}{c|}{86.4}  & 7.2     & 0.1     & 80.1    & 87.4  \\
\multicolumn{1}{c|}{$y_{i4}$} & \multicolumn{1}{c|}{0.5}     & 7.0     & 79.5    & \multicolumn{1}{c|}{86.5}  & 7.0     & 0.0     & 80.6    & 87.6  \\
\multicolumn{1}{c|}{$y_{i5}$} & \multicolumn{1}{c|}{84.3}    & 0.2     & 0.3     & \multicolumn{1}{c|}{0.6}   & 0.1     & 88.1    & 0.1     & 88.4  \\
\multicolumn{1}{c|}{$y_{i6}$} & \multicolumn{1}{c|}{81.2}    & 0.1     & 1.1     & \multicolumn{1}{c|}{1.2}   & 1.4     & 85.9    & 0.9     & 99.2  \\ \hline
\end{tabular}
\end{table}

\clearpage

\section{Applications to real data} \label{sec4}

In this section, the quantile factor model, Normal and Student-t factor models are evaluated using two real datasets with two very different structures. In the first application, 5 daily price indexes are analyzed, for which the correlation structure seems to be highly dependent on the quantile considered, motivating the use of our proposed. The second application is to a dataset from a heart disease study. In this case, the correlation structure a priori does not seem to be quantile dependent, not favoring the use of the proposed model.

For both cases, we considered the same hyperparameters described in Section \ref{sec3} for the prior distributions. The MCMC algorithm was also implemented in the \texttt{R} programming language, version 3.4.1 \cite{teamR}, in a computer with an Intel(R) Core(TM) 
i5-4590 processor with 3.30 GHz and 8 GB of RAM memory. For each sample and fitted model, we also ran two parallel chains starting from different values, letting each chain run for 160,000 iterations, discarded the first 10,000 as burn-in, and stored every 50th iteration to avoid possible autocorrelations within the chains. We also used the diagnostic tools available in the CODA package \cite{plummer2006coda} to check convergence of the chains.

\subsection{Financial sector dataset}\label{sec:aplicacao3-realdata-1-setorfinanc}

The dataset consists of 5 daily price indexes during 2018. In particular, variables $y_{i1}$ to $y_{i5}$ for $i = 1, \dots, n$ represent, respectively the Russell 2000 (New York, USA), Bell 20 (Brussels, Belgium), IBEX 35 (Madrid, Spain), Straits Times (Singapore) and Hang Seng (Hong Kong, China) indexes. This dataset was also analyzed to illustrate the quantile correlation measure in \cite{choi2018quantile} and is available at \url{https://realized.oxford-man.ox.ac.uk/}

\begin{figure}[h!]
    \centering
    \includegraphics[scale = 0.65]{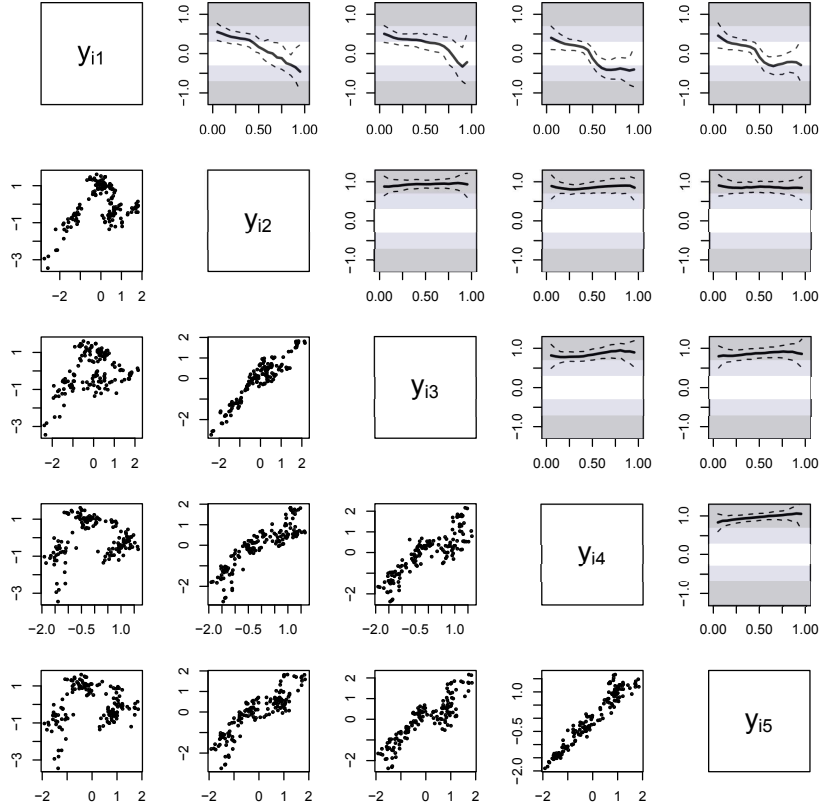}
    \caption{Lower panels: Scatterplots of the price indexes. Upper panels: posterior mean (solid line) with respective 95\% credible interval (dashed line) of the pairwise quantile correlation varying by quantile. The gray scale highlights regions for which the correlation is weak ($|\rho_\tau| < 0.3$), moderate ($ 0.3 < |\rho_\tau| < 0.7$) and strong ($|\rho_\tau| > 0.7$).}
    \label{fig:pairs-y-realdata}
\end{figure}

Figure \ref{fig:pairs-y-realdata} shows that $y_{i2}$, $y_{i3}$, $y_{i4}$ are $y_{i5}$ are correlated for all the quantiles, but the quantile correlation between the Russell 2000 index ($y_{i1}$) and the other indexes is 
weak in the intermediate quantiles and moderate in the extreme quantiles, and also is positive in the lowest quantiles, and negative in the highest ones.

\begin{figure}[h!]
    \centering
    \includegraphics[scale = 0.6]{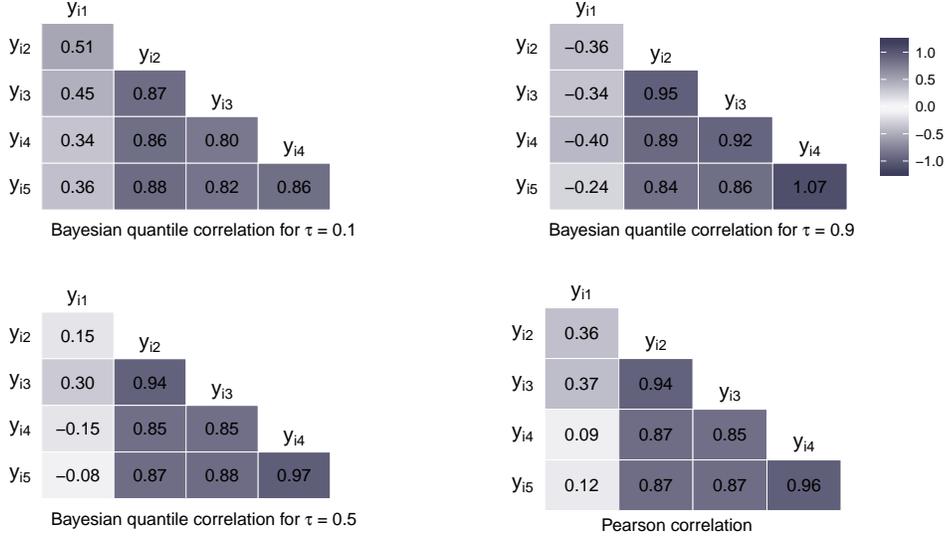}
    \caption{Quantile correlation matrix estimated for $\tau=0.1, 0.5$ and $0.9$ and Pearson correlation matrix for the financial index dataset.}
    \label{fig:matcor-realdata}
\end{figure}

Figure \ref{fig:matcor-realdata} displays the quantile correlation matrices for $\tau = 0.1, 0.5 \textrm{ and } 0.9$ and the linear correlation matrix. Similar conclusions are obtained in this case.  The linear correlation matrix is similar to the quantile correlations for all the indexes, except the Russell 2000 ($y_{i1}$), for which the highest values for the quantile correlation is obtained when $\tau = 0.1$. It is interesting to observe that for $\tau=0.9$ the quantile correlations between the Russell 2000 and the other indexes are all negative.

Table \ref{table-medidas-resumo-realdata} reports the comparison criteria values in the quantile factor model for some values of $\tau$, and the Student-t and Normal factor model fits. When $k=2$ is assumed, RPS, MAE and MSE point to the quantile factor model as the best, especially with $\tau = 0.1$ and $0.9$.

On the other hand, ICOMP, AIC, BIC and BIC* suggest that quantile factor models assuming $k=2$ always perform best for all the quantiles considered, while for the Student-t and Normal factor model fits they suggest $k=1$. 

\begin{table}[h!]
\centering
\caption{Model comparison criteria for the quantile, Normal and Student-t factor models assuming $\tau = 0.1, 0.25, 0.5, 0.75 \textrm{ and } 0.9$, $k=1$ and $k=2$.}
\label{table-medidas-resumo-realdata}\vspace{0.2 cm}
\begin{tabular}{lcccccccc}
\hline
Model & $k$ & ICOMP            & AIC              & BIC              & BIC*             & RPS           & MAE           & MSE           \\ \hline
$\textrm{QFM}_{\tau = 0.10}$   & 1   & 1334.66          & 1350.15          & 1380.26          & 1379.98          & 0.33          & 0.42          & 0.55          \\ \hline
$\textrm{QFM}_{\tau = 0.25}$   & 1   & 1236.79          & 1251.15          & 1281.26          & 1280.97          & 0.30          & 0.39          & 0.41          \\ \hline
$\textrm{QFM}_{\tau = 0.50}$   & 1   & 1168.63          & 1181.54          & 1211.65          & 1211.37          & 0.26          & 0.36          & 0.28          \\ \hline
$\textrm{QFM}_{\tau = 0.75}$   & 1   & 1256.85          & 1271.38          & 1301.49          & 1301.21          & 0.28          & 0.36          & 0.34          \\ \hline
$\textrm{QFM}_{\tau = 0.90}$   & 1   & 1371.69          & 1387.18          & 1417.29          & 1417.01          & 0.32          & 0.40          & 0.46          \\ \hline
TFM     & 1   & 853.85            & 840.63            & 810.52            & 810.80            & 0.26          & 0.35          & 0.28          \\ \hline
NFM     & 1   & 14.20            & 28.58            & 58.68            & 58.40            & 0.26          & 0.35          & 0.27          \\ \hline
$\textrm{QFM}_{\tau = 0.10}$   & 2   & 1018.54 & 1045.98 & 1088.13 & 1087.67 & 0.14 & 0.16 & 0.05 \\ \hline
$\textrm{QFM}_{\tau = 0.25}$   & 2   & 1011.01 & 1037.61 & 1079.76 & 1079.30 & 0.16          & 0.18          & 0.06          \\ \hline
$\textrm{QFM}_{\tau = 0.50}$   & 2   & 1031.39          & 1051.75          & 1093.90          & 1093.44          & 0.18          & 0.20          & 0.09          \\ \hline
$\textrm{QFM}_{\tau = 0.75}$   & 2   & 1053.68          & 1079.38          & 1121.53          & 1121.07          & 0.17          & 0.18          & 0.06          \\ \hline
$\textrm{QFM}_{\tau = 0.90}$   & 2   & 1096.82          & 1124.38          & 1166.53          & 1166.07          & 0.15          & 0.17          & 0.05          \\ \hline

TFM     & 2   & 1443.50            & 1424.90            & 1382.27            & 1383.21            & 0.18         & 0.22          & 0.11          \\ \hline

NFM     & 2   & 15.09            & 35.51            & 77.66            & 77.20            & 0.19          & 0.22          & 0.11          \\ \hline
\end{tabular}
\end{table}

\begin{figure}[h!]
    \centering
    \subfigure[$k=1$]{\includegraphics[scale = 0.22]{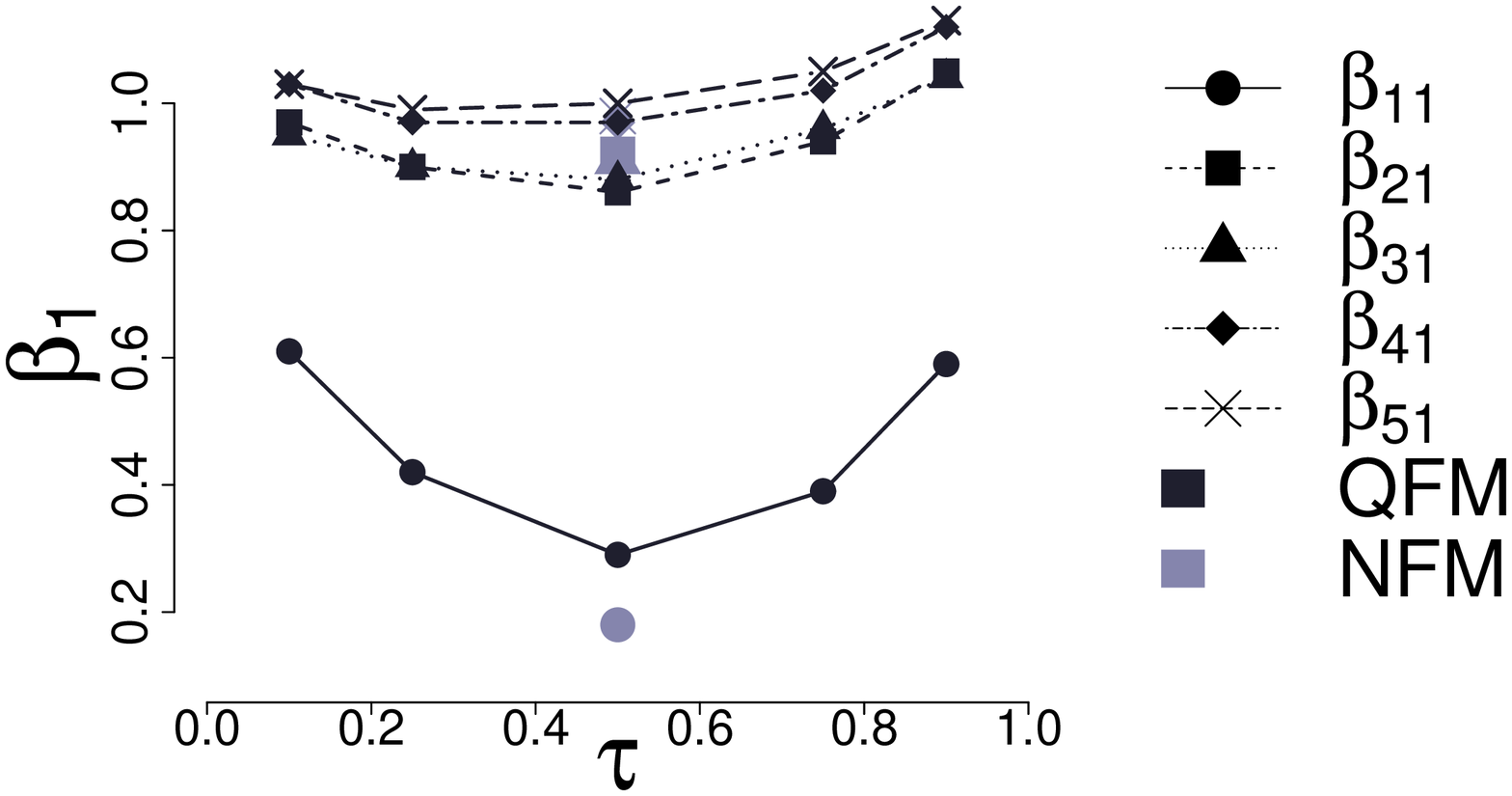}}
    \subfigure[$k=2$]{\includegraphics[scale = 0.27]{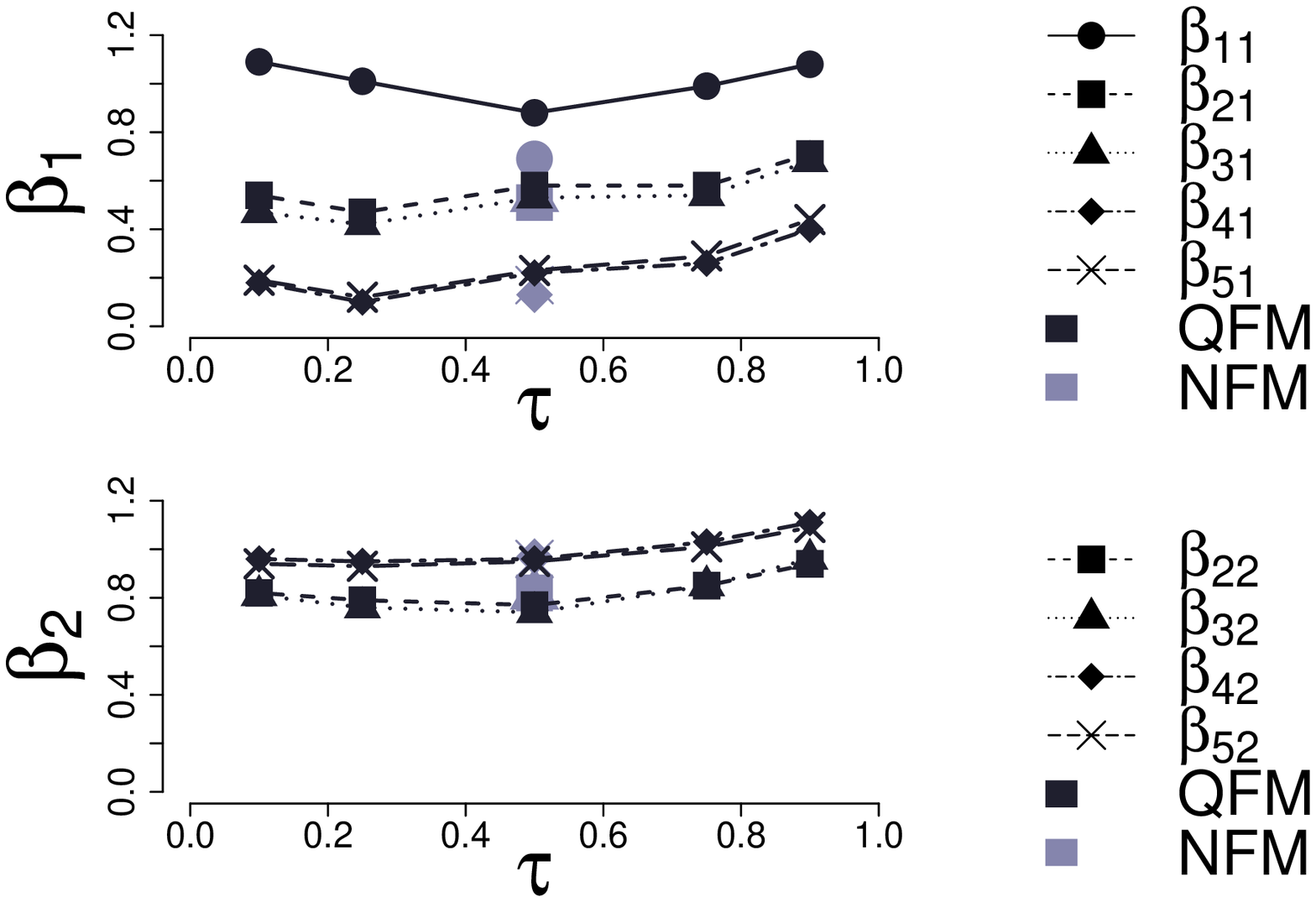}}
    \caption{Posterior mean of the factor loadings matrix obtained in the quantile factor model fit assuming $\tau = 0.1, 0.25, 0.5, 0.75 \textrm{ and } 0.9$, $k=1$ and $k=2$ and the Normal factor model.}
    \label{fig:mat-beta-dadosreais-k1}
\end{figure}

Figures \ref{fig:mat-beta-dadosreais-k1} (a) and \ref{fig:mat-beta-dadosreais-k1} (b) show the posterior mean of the factor loadings obtained under the quantile (in black) and Normal (in gray) factor model fits assuming $k=1$ and 2. When $k = 1$, the latent factor is strongly correlated to variables $y_{i2}, y_{i3}, y_{i4} \textrm{ and } y_{i5}$, as expected. Although the latent factor does not explain well the Russell 2000 index ($y_{i1}$) in general, some improvement can be viewed in the factor loadings for extreme quantiles. 
On the other hand, from Figure \ref{fig:mat-beta-dadosreais-k1} (b), we notice that while the second latent factor is correlated to variables $y_{i2}, y_{i3}, y_{i4} \textrm{ and } y_{i5}$ for both models and all the quantiles, the first latent factor is more correlated to the Russell 2000 index ($y_{i1}$), mainly in 10\% and 90\% quantiles, although it seems to be also correlated to the Bell 20 and IBEX 35 indexes ($y_{i2} \textrm{ and } y_{i3}$). The Normal factor model does not capture well the Russell 2000 index in its structure.

Moreover, from Table \ref{dv-realdata} with the variance decomposition for each model considered, we conclude that while the variability explained by the latent factor for the Russell 2000 index ($y_{i1}$) in the Normal and Student-t model fits does not exceed 50\%, it is higher than 75\% with just one factor for $\tau = 0.1$  and $0.9$. In the two-quantile factor model, all the indexes are well explained, mainly for $\tau = 0.1$  and $0.9$, for which the first latent factor is clearly related to $y_{i1}$ and the second latent factor to the other components. The modified variance decomposition in equation (\ref{DVmod}) is presented just to reinforce the previous conclusion.

\begin{table}[h!]
\centering
\caption{Posterior mean of the modified variance decomposition and variance decomposition obtained, respectively, in the quantile factor model fit with $\tau=0.1$, 0.5 and 0.9, and the Normal and Student-t factor model fits, assuming $k=1$ and $2$.\label{dv-realdata}}\vspace{0.3 cm}
\begin{tabular}{c|c|ccc|c|ccc} 
\hline
\multicolumn{1}{c}{}           & \multicolumn{4}{c|}{Variance decomposition} & \multicolumn{4}{c}{Modified variance decomposition}      \\ 
\cline{2-9}
\multicolumn{1}{c}{}           & k = 1    & \multicolumn{3}{c|}{k = 2}       & \multicolumn{1}{c}{k = 1} & \multicolumn{3}{c}{k = 2}    \\ 
\cline{2-9}
\multicolumn{1}{c}{}           & Fac 1 & Fac 1 & Fac 2 & Total       & Fac 1                  & Fac 1 & Fac 2 & Total  \\ 
\hline
\multicolumn{1}{l}{}           & \multicolumn{8}{c}{$\textrm{QFM}_{\tau = 0.1}$ }                                                       \\ 
\hline
$y_{i1}$                       & 12.8     & 94.8    & 0.0      & 94.8        & 76.6                      & 99.8     & 0.0      & 99.8   \\
$y_{i2}$                       & 74.3     & 26.1    & 61.2     & 87.3        & 98.5                      & 29.7     & 69.6     & 99.4   \\
$y_{i3}$                       & 73.3     & 21.9    & 65.0     & 86.9        & 98.4                      & 25.1     & 74.3     & 99.3   \\
$y_{i4}$                       & 76.6     & 3.0     & 87.8     & 90.8        & 98.6                      & 3.2      & 96.3     & 99.5   \\
$y_{i5}$                       & 77.0     & 3.7     & 86.8     & 90.5        & 98.7                      & 4.1      & 95.5     & 99.5   \\ 
\hline
\multicolumn{1}{l}{}           & \multicolumn{8}{c}{$\textrm{QFM}_{\tau = 0.5}$ }                                                       \\ 
\hline
\multicolumn{1}{l|}{$y_{i1}$ } & 5.7      & 56.8    & 0.0      & 56.8        & 32.5                      & 91.3     & 0.0      & 91.3   \\
\multicolumn{1}{l|}{$y_{i2}$ } & 76.6     & 33.8    & 57.9     & 91.6        & 96.3                      & 36.4     & 62.4     & 98.9   \\
\multicolumn{1}{l|}{$y_{i3}$ } & 75.1     & 29.8    & 54.4     & 87.2        & 96.0                      & 33.6     & 94.6     & 98.2   \\
$y_{i4}$                       & 93.4     & 4.8     & 90.7     & 95.5        & 99.1                      & 5.0      & 94.4     & 99.4~  \\
$y_{i5}$                       & 95.3     & 5.2     & 89.7     & 94.9~       & 99.4                      & 5.4      & 93.9     & 99.3~  \\ 
\hline
\multicolumn{1}{c}{}           & \multicolumn{8}{c}{$\textrm{QFM}_{\tau = 0.9}$ }                                                       \\ 
\hline
$y_{i1}$                       & 12.0     & 89.8    & 0.0      & 89.8        & 75.1                      & 99.5     & 0.0      & 99.5   \\
$y_{i2}$                       & 77.1     & 32.2    & 58.4     & 90.6        & 98.7                      & 35.3     & 64.2     & 99.5   \\
$y_{i3}$                       & 75.7     & 29.5    & 61.0     & 90.4        & 98.6                      & 32.4     & 67.1     & 99.5   \\
$y_{i4}$                       & 79.1     & 10.4    & 83.9     & 93.3        & 98.8                      & 11.0     & 88.7     & 99.7   \\
$y_{i5}$                       & 80.1     & 12.3    & 81.8     & 94.0        & 98.9                      & 13.0     & 86.7     & 99.7   \\ 
\hline
\multicolumn{9}{c}{Variance decomposition}                                                                                              \\ 
\hline
\multicolumn{1}{c}{}           & \multicolumn{4}{c|}{NFM}                    & \multicolumn{4}{c}{TFM}                                  \\ 
\hline
$y_{i1}$                       & 3.3      & 46.8    & 0.0      & 46.8        & 0.1                       & 43.8     & 0.0      & 43.8   \\
$y_{i2}$                       & 82.8     & 25.9    & 66.4     & 92.3        & 80.9                      & 29.5     & 63.9     & 93.4   \\
$y_{i3}$                       & 81.7     & 27.7    & 64.6     & 92.3        & 80.0                      & 31.0     & 62.1     & 93.1   \\
$y_{i4}$                       & 93.2     & 1.8     & 93.3     & 95.1        & 94.8                      & 3.0      & 93.3     & 96.3   \\
$y_{i5}$                       & 94.1     & 2.8     & 92.1     & 94.9        & 96.2                      & 4.2      & 91.7     & 95.9   \\
\hline
\end{tabular}
\end{table}


\subsection{Heart disease dataset}\label{sec:aplicacao4-realdata-wcgs}

Here we analyze a sample extracted from the dataset {\tt wcgs} available in the \textit{faraway} package of R \cite{citation_faraway}. It consists of data on
3154 healthy young men aged from 39 to 59 years old from the San Francisco area. All patients were free from coronary heart disease at the start of the study. Eight and a half years later, changes in this situation were recorded. In particular, this application was performed with a random sample of $n=100$ individuals and is concentrated in the following variables: weight ($y_{i1}$), height ($y_{i2}$), systolic blood pressure ($y_{i3}$), diastolic blood pressure ($y_{i4}$) and fasting serum cholesterol ($y_{i5}$). 

\begin{figure}[h!]
    \centering
    \includegraphics[scale = 0.6]{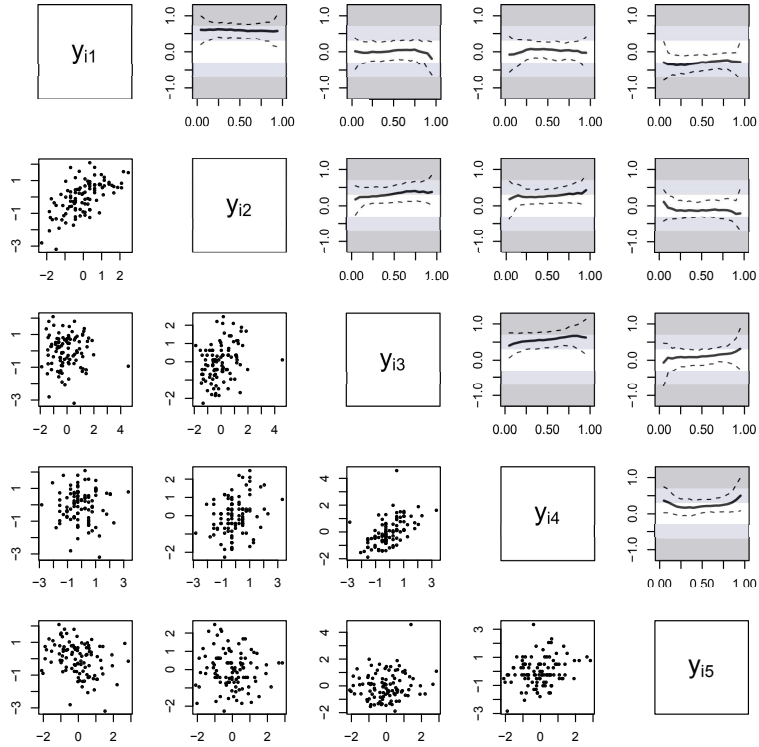}
    \caption{Lower panels: Scatterplots of the heart disease dataset. Upper panels: posterior mean (solid line) with respective 95\% credible interval (dashed line) of the pairwise quantile correlation varying by quantile. The gray scale highlights regions for which the correlation is weak ($|\rho_\tau| < 0.3$), moderate ($ 0.3 < |\rho_\tau| < 0.7$) and strong ($|\rho_\tau| > 0.7$).}
    \label{fig:pairs-y-realdata2}
\end{figure}

From Figure \ref{fig:pairs-y-realdata2} it can be seen that weight and height ($y_{i1}$ and $y_{i2}$, respectively) are strongly correlated, as are the systolic and diastolic blood pressure ($y_{i3}$ and $y_{i4}$, respectively). The aim of this application is to evaluate the proposed quantile model's performance in comparison with the Student-t and the Normal ones, in a real scenario that does not seem to favor a quantile analysis a priori.

\begin{figure}[h!]
    \centering
    \includegraphics[scale = 0.6]{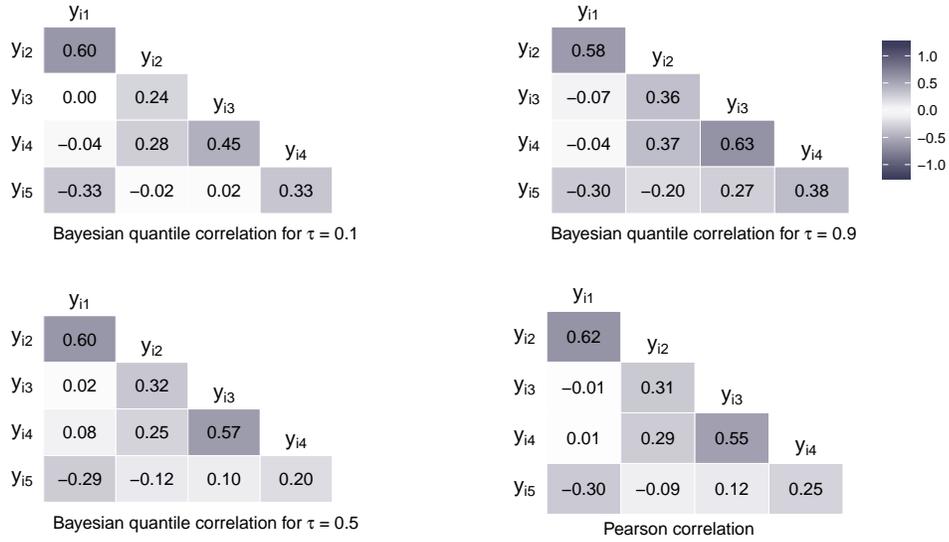}
    \caption{Quantile correlation matrix estimate for $\tau=0.1, 0.5$ and $0.9$ and Pearson correlation matrix for the heart disease dataset.}
    \label{fig:matcor-realdata2}
\end{figure}

Additionally, Figure \ref{fig:matcor-realdata2} presents the quantile correlation matrix for $\tau = 0.1, 0.5 \textrm{ and } 0,9$ and the linear correlation. In fact, the quantile correlation is very similar to the linear one, mainly when $\tau=0.5$ is assumed. Thus, in this application we just considered the proposed quantile factor model with $\tau=0.5$ and compared it to the Normal and Student-t factor model fits, assuming $k=1$ and $k=2$.

Table \ref{table-medidas-resumo-realdata2} reports the model comparison criteria results for the three models fitted. First, with respect to the number of factors, while all the criteria point to $k=2$ as the best setting in the quantile and Student-t factor models, AIC, BIC and BIC* point to $k=1$ as the best in the Normal fit. Then, for $k = 2$ fixed, RPS, MAE and MSE criteria do not differ much across the different models. However, while the RPS criterion suggests the quantile factor model in the median performs best, MAE suggests the Student-t factor model and MSE equally suggests the Student-t and Normal factor models as the best.

\begin{table}[h!]
\centering
\caption{Model comparison criteria for the quantile factor model for $\tau=0.5$, and the Normal and Student-t factor models for $k=1$ and $k=2$.}
\label{table-medidas-resumo-realdata2}\vspace{0.3 cm}
\begin{tabular}{lcccccccc}
\hline
Model             & k & ICOMP & AIC & BIC & BIC* & RPS & MAE & MSE \\ \hline
$\textrm{QFM}_{\tau = 0.5}$ & 1          & 1259.28        & 1276.97      & 1303.02      & 1302.59       & 0.43         & 0.58         & 0.66         \\ \hline
NFM                          & 1          & 14.70          & 33.51        & 59.57        & 59.14         & 0.45         & 0.59         & 0.65         \\ \hline
TFM                          & 1          & 243.45         & 261.86       & 287.91       & 287.48        & 0.44         & 0.57         & 0.64         \\ \hline
$\textrm{QFM}_{\tau = 0.5}$ & 2          & 1222.10        & 1247.25      & 1283.72      & 1283.03       & 0.35         & 0.42         & 0.38         \\ \hline
NFM                          & 2          & 13.99          & 40.96        & 77.44        & 76.74         & 0.37         & 0.43         & 0.34         \\ \hline
TFM                          & 2          & 7.85           & 34.58        & 71.06        & 70.36         & 0.36         & 0.41         & 0.34\\\hline        
\end{tabular}
\end{table}

Table \ref{matriz-beta-realdata2} reports the posterior mean of the factor loadings for the models considered. The results are, in general, very similar. When $k = 1$, the latent is mainly related to weight and height ($y_{i1}$ and $y_{i2}$), while fasting serum cholesterol ($y_{i5}$) presents the lowest value of the associated factor loading, reflecting the structure presented in Figures \ref{fig:pairs-y-realdata2} and \ref{fig:matcor-realdata2}. On the other hand, when $k=2$ is fixed, there is clear separation of the original variables among the two latent factors considered. While, the first factor is related to weight and height ($y_{i1}$ and $y_{i2}$), the second factor explains the systolic and  diastolic blood pressure ($y_{i3}$ and $y_{i4}$). The fasting serum cholesterol is almost equally explained by both factors. Moreover, since Figures \ref{fig:pairs-y-realdata2} and \ref{fig:matcor-realdata2} show a possible inverse correlation between cholesterol component and weight and height components, which are well explained by the first factor, we would expect a negative value for the factor loading of this component. However, this happens only for the quantile factor model in the median and the Normal one. In general the results are very similar for the three models fitted.

\begin{table}[h!]
\centering
\caption{Posterior mean of the factor loadings matrix obtained in the quantile factor model fit assuming $\tau =0.5$, and the Normal and Student-t factor models for $k=1$ and $2$.\label{matriz-beta-realdata2}}\vspace{0.3 cm}
\begin{tabular}{c|c|c|cc|cc|cc}
\hline
\multicolumn{3}{c|}{$k=1$}                   & \multicolumn{6}{c}{$k=2$}                                                                             \\ \hline
$\textrm{QFM}_{\tau = 0,5}$ & NFM    & TFM   & \multicolumn{2}{c|}{$\textrm{QFM}_{\tau = 0.5}$} & \multicolumn{2}{c|}{NFM} & \multicolumn{2}{c}{TFM} \\ \hline
0.57                        & 0.68  & 0.59 & 0.66                    & 0.00                   & 0.85        & 0.00      & 0.80       & 0.00       \\
0.97                        & 0.90  & 0.86 & 0.90                    & 0.30                   & 0.73        & 0.40      & 0.69       & 0.36       \\
0.24                        & 0.30  & 0.34 & 0.06                    & 0.64                   & 0.02        & 0.72      & 0.05       & 0.63       \\
0.29                        & 0.29  & 0.31 & 0.05                    & 0.88                   & 0.00        & 0.76      & 0.01       & 0.76       \\
-0.16                       & -0.14 & 0.12 & -0.20                   & 0.16                   & -0.32       & 0.27      & 0.30       & 0.21  \\ \hline    
\end{tabular}
\end{table}

Finally, Table \ref{dv-realdata2} presents the posterior mean of the variance decomposition for the three models considered. In general, the two-factor structure well explains the original variability for the three models considered, except for the fasting serum cholesterol component ($y_{i5}$). The Normal and Student-t factor models present similar results. On the other hand, for comparison purposes, these results reveal a case for which the modified variance decomposition in equation (\ref{DVmod}) may be useful. For example, when $k=1$, although the latent factor is strongly related to weight and height components, the proposed model better explains the the systolic and  diastolic blood pressure than the other models according to the modified variance decomposition. Moreover, under the variance decomposition the quantile factor model evaluated at the median explains about 40\% of the variability assuming the two-factor structure, while under the modified variance decomposition this percentage increases to 80\%.

\begin{table}[h!]
\centering
\caption{Posterior mean of the modified variance decomposition and variance decomposition obtained in the quantile factor model fit with $\tau=0.5$, and Normal and Student-t factor model fits, assuming $k=1$ and 2.\label{dv-realdata2}}\vspace{0.3 cm}
\begin{tabular}{ccccccccc}
\hline
                           & \multicolumn{8}{c}{$\textrm{QFM}_{\tau = 0.5}$}                                                                                       \\ \cline{2-9} 
                           & \multicolumn{1}{c|}{k = 1}  & \multicolumn{3}{c|}{k = 2}                   & \multicolumn{1}{c|}{k = 1}   & \multicolumn{3}{c}{k = 2} \\ \cline{2-9} 
                           & \multicolumn{1}{c|}{Fac 1} & Fac 1 & Fac 2 & \multicolumn{1}{c|}{Total} & \multicolumn{1}{c|}{Fac 1}  & Fac 1  & Fac 2  & Total \\ \cline{2-9} 
                           & \multicolumn{4}{c|}{Variance decomposition}                             & \multicolumn{4}{c}{Modified variance decomposition} \\ \hline
\multicolumn{1}{c|}{$y_{i1}$} & \multicolumn{1}{c|}{28.5}   & 38.8   & 0.0    & \multicolumn{1}{c|}{38.8}  & \multicolumn{1}{c|}{76.1}    & 83.6    & 0.0     & 83.6  \\
\multicolumn{1}{c|}{$y_{i2}$} & \multicolumn{1}{c|}{87.6}   & 74.5   & 8.5    & \multicolumn{1}{c|}{83.0}  & \multicolumn{1}{c|}{98.3}    & 87.5    & 10.0    & 97.5  \\
\multicolumn{1}{c|}{$y_{i3}$} & \multicolumn{1}{c|}{5.4}    & 0.3    & 40.7   & \multicolumn{1}{c|}{41.0}  & \multicolumn{1}{c|}{31.4}    & 0.7     & 84.1    & 84.7  \\
\multicolumn{1}{c|}{$y_{i4}$} & \multicolumn{1}{c|}{7.4}    & 0.2    & 71.4   & \multicolumn{1}{c|}{71.6}  & \multicolumn{1}{c|}{38.9}    & 0.3     & 95.0    & 95.3  \\
\multicolumn{1}{c|}{$y_{i5}$} & \multicolumn{1}{c|}{2.0}    & 3.2    & 2.1    & \multicolumn{1}{c|}{5.3}   & \multicolumn{1}{c|}{14.1}    & 18.7    & 12.3    & 30.9  \\ \hline
\multicolumn{1}{l}{}       & \multicolumn{4}{c|}{NFM}                                                   & \multicolumn{4}{c}{TFM}                                  \\ \cline{2-9} 
\multicolumn{1}{l}{}       & \multicolumn{4}{c|}{Variance decomposition}                             & \multicolumn{4}{c}{Variance decomposition}            \\ \hline
\multicolumn{1}{l|}{$y_{i1}$} & \multicolumn{1}{c|}{45.2}   & 70.1   & 0.0    & \multicolumn{1}{c|}{70.1}  & \multicolumn{1}{c|}{38.4}    & 70.0    & 0.0     & 70.0  \\
\multicolumn{1}{l|}{$y_{i2}$} & \multicolumn{1}{c|}{77.0}   & 52.2   & 15.2   & \multicolumn{1}{c|}{67.4}  & \multicolumn{1}{c|}{82.6}    & 50.3    & 13.4    & 63.6  \\
\multicolumn{1}{l|}{$y_{i3}$} & \multicolumn{1}{c|}{8.9}    & 0.0    & 51.1   & \multicolumn{1}{c|}{51.1}  & \multicolumn{1}{c|}{11.4}    & 0.3     & 47.1    & 47.4  \\
\multicolumn{1}{c|}{$y_{i4}$} & \multicolumn{1}{c|}{8.0}    & 0.0    & 56.8   & \multicolumn{1}{c|}{56.8}  & \multicolumn{1}{c|}{8.8}     & 0.0     & 64.4    & 64.4  \\
\multicolumn{1}{c|}{$y_{i5}$} & \multicolumn{1}{c|}{1.9}    & 9.7    & 7.7    & \multicolumn{1}{c|}{17.4}  & \multicolumn{1}{c|}{1.6}     & 9.6    & 4.6     & 14.2  \\ \hline
\end{tabular}
\end{table}

\section{Conclusions} \label{sec5}

We propose a new class of models, named quantile factor models, which is advantageous compared to standard factor models since it provides richer information about the latent factors, is
robust to heteroscedasticity and outliers, and can accommodate the non-normal errors often encountered in practical applications. The method not only allows factor loadings to vary across the quantiles, but also the latent factors. For the inference procedure, we presented a MCMC algorithm based mostly
on Gibbs sampling for the location-scale mixture representation of the $\mathcal{AL}_p$ distribution.
The method is also an alternative to the quantile correlation, which may be used in some applications as a preliminary exploratory measure. 

We evaluated the QFM in artificial and real datasets and compared it with the Normal and Student-t factor models. We concluded that the proposed model is a robust alternative to these, in some cases having similar performance to the Student-t one. On the other hand, the flexibility of the method shows that the quantile to be tracked depends on the specific aims of the problem and different results and interpretations can be obtained for each case. Model comparison criteria and variance decomposition were used not only to evaluate the model's performance under different quantiles but also to infer about latent factor dimensions for each quantile considered.

\appendix

\section{Multivariate asymmetric Laplace distribution}

A random vector $X \in \mathbb{R}^p$ is said to have ($\mathcal{AL}_p$) distribution with parameters $\mu$ and $\Psi$, this is, $X\sim \mathcal{AL}_p(\mu, \Psi)$ if its characteristic function is given by:
\begin{equation*}
    \psi(t) = \frac{1}{1 + \frac{1}{2}t'\Psi t - i\mu't},
\end{equation*}
where $m \in \mathbb{R}^p$ and $\psi$ is a $p \times p$ non-negative definite symmetric matrix \cite[ch. 6]{kotz2012laplace}. Moreover, we have that $E(\textbf{X}) = \mu$ and $Var(\textbf{X}) = \mu\mu' + \Psi.$

As in the univariate case, the $\mathcal{AL}_p$ distribution admits a location-scale mixture representation of Normal and Exponential distributions. Let $V \sim N_p(0, \Psi)$ and $W \sim Exp(1)$, then we can write $X$ using the following mixture representation:
\begin{equation}\label{eq:mistura-laplace-assimetrica-multivariada}
    X = \mu W  + \sqrt{W} V. 
\end{equation}

Moreover, all univariate marginals are $\mathcal{LA}$-distributed. That is, for $X = (X_1, \dots, X_p)'$, $\mu = (\mu_1, ..., \mu_p)'$ and $\Psi=(\psi_{ij})_{i,j=1}^p$, we have that $X_{l} \sim \mathcal{AL}(\mu_l, \psi_l)$ for all $l = 1, \dots, p$ and:
\begin{equation*}
E(X_{l}) = \mu_l ,\; Var(X_{l}) = \mu_l^2+\psi_{ll} \, \mbox { and } \,	Cov(X_{l}, X_{h}) =  \mu_l\mu_{h} + \psi_{lh}.
\end{equation*}

Coming back to the notation used in the proposed model in (\ref{eq:definicao-mfq}), we get: $\mu=m$, $\Psi=\Delta$. On the other hand, the univariate quantile regression in (\ref{eq:mistura-laplace-assimetrica-univariada}) is obtained in this case assuming that $\mu_l= m_l= \sigma_{l} a_\tau $ and $\psi_{ll}= \delta_{ll} = \sigma^2_{l} b_\tau^2$, for $a_\tau$ and $b_\tau^2$ defined in equation (\ref{eq:mistura-laplace-assimetrica-univariada}). 

More details about the $\mathcal{AL}_p$ distribution can be seen in \cite[ch. 6]{kotz2012laplace}.

\section{Full conditional posterior distributions of the parameters in the proposed model}

In this section we present the posterior full conditional distributions of the components of the parameter vector $\Theta$. We denote
the posterior full conditional of a parameter $\theta$ in $\Theta$ by $p(\theta\mid \dots)$.

\subsection{Full conditional posterior distribution of \texorpdfstring{$w_i$}:}

For $i = 1, \dots, n$, the posterior full conditional of the location-scale mixture parameter $w_i$ is proportional to:

\begin{eqnarray*}
p(w_i\mid \dots) &\propto &  \displaystyle  \prod_{i=1}^{n} p(y_i | f_i, w_i)  p(f_i) p(w_i)\\
& \propto & w_i^{(1-p)/2}\exp\left\{(2 + m'\Delta^{-1}m)w_i+(y_i - \beta f_i)'\Delta^{-1}(y_i - \beta f_i)w_i^{-1}\right\} 
\end{eqnarray*}

Therefore,
     \begin{equation*}
        w_i \mid \cdot \sim \mathcal{GIG}\left(1-\frac{p}{2}, 2 + m'\Delta^{-1}m, (y_i - \beta f_i)'\Delta^{-1}(y_i - \beta f_i) \right) ,
    \end{equation*}
    where $\mathcal{GIG}$ denotes the generalized inverse Gaussian distribution \cite{jorgensen2012statistical}.

\subsection{Full conditional posterior distribution of \texorpdfstring{$f_i$}:}
    For $i = 1, ..., n$, the posterior full conditional of the latent factors $f_i$ is proportional to:
\begin{eqnarray*}
p(f_i\mid \dots) &\propto &   \displaystyle  \prod_{i=1}^{n} p(y_i | f_i, w_i)  p(f_i) \\
& \propto & \exp \left[-\frac{1}{2} \left(\displaystyle \sum_{i=1}^{n}(y_i - \beta f_i - mw_i)'(w_i \Delta)^{-1}(y_i - \beta f_i - mw_i) - \displaystyle \sum_{i=1}^{n}f_i'I_kf_i  \right)\right]
\end{eqnarray*}    
Therefore,
    \begin{equation*}
       f_i \mid \cdot \sim N\displaystyle \left( \Omega \beta'(w_i \Delta)^{-1}(y_i - m w_i), \Omega \right), \quad \textrm{where} \quad \Omega^{-1} = \beta'(w_i \Delta)^{-1}\beta + I_k  
    \end{equation*}

\subsection{Full conditional posterior distribution of \texorpdfstring{$\beta_j$}:}
    Recall that $\beta_j$ denotes the $j$-th column of factor loadings matrix $\beta$. For $j = 1, \dots, p$, the posterior full conditional of $\beta_j$ is proportional to:    
    
    \begin{eqnarray*}
p(\beta_j\mid \dots) &\propto &   \displaystyle  \prod_{i=1}^{n} p(y_i | \beta, f_i, w_i)  	\displaystyle  \prod_{j=1}^{k} p(\beta_j) \displaystyle  \prod_{j=k+1}^{p} p(\beta_j) \\
& \propto & \exp \left(-\frac{1}{2} \displaystyle \sum_{i=1}^{n}(y_i - \beta f_i - mw_i)'(w_i \Delta)^{-1}(y_i - \beta f_i - mw_i)\right)\\
& \times & \displaystyle \prod_{j=1}^k   \textrm{exp} \left( -\frac{1}{2} \beta_j (C_0  I_j)^{-1} {\beta_j}'  \mathbb{I}(\beta_{jj} \geq 0) \right)   \displaystyle \prod_{j=k+1}^p  \textrm{exp} \left( -\frac{1}{2} \beta_j (C_0I_k)^{-1} {\beta_j}' \right) 
\end{eqnarray*}

    Therefore,
    \begin{itemize}
     \item for $j \leq k$:
     \begin{align*}
    & \beta_j | \cdot \sim N(C_1^* r^*, C_1^*) \mathbb{I}(\beta_{jj}>0), \quad \textrm{where} \\ & \quad {C_1^*}^{-1} = C_0 I_j + \displaystyle \sum_{i=1}^{n} w_i^{-1} \delta_{jj}^{-2} f_i^j {f_i^j}' \quad \textrm{and} \quad r^* = \displaystyle \sum_{i=1}^{n} f_i^j \delta_{jj}^{-2} (w_i^{-1} y_{ij} - m_j)   
    \end{align*}
    
    \item for $j > k$:
    \begin{align*}
   & \beta_j | \cdot \sim  N(C_1 r, C_1), \quad \textrm{ where } \\ & \quad C_1^{-1} = C_0 I_k + \displaystyle \sum_{i=1}^{n} w_i^{-1} \delta_{jj}^{-2} f_i f'_i \quad \textrm{e} \quad r = \displaystyle \sum_{i=1}^{n} f_i \delta_{jj}^{-2} (w_i^{-1} y_{ij} - m_j)    
    \end{align*}
    
    \end{itemize}
    
\subsection{Full conditional posterior distribution of \texorpdfstring{$\sigma^{-2}_j$}:}    
For $j=1,\dots,p$, the posterior full conditional of the inverse of idiosyncratic
variances  $\sigma^{-2}_j$ is proportional to:  
        \begin{equation*}
	\begin{array}{lll}
	p(\sigma^{-2}_j | \dots ) \propto & 
(\sigma^{-2}_j)^{\frac{n + \nu}{2}-1} \exp \left\{ (-\sigma^{-2}_j) \left( \frac{\nu s^2}{2} +  \frac{\tau(1-\tau)}{4} \displaystyle \sum_{i=1}^n \frac{(y_{ij} - \beta_j f_i)^2}{w_i}  \right) \right\}\\
	& \exp \left\{ (-\sigma^{-1}_j) \left( \frac{1-2\tau}{2} \displaystyle \sum_{i=1}^n (y_{ij} - \beta_j f_i) \right)  \right\},
	\end{array}
	\end{equation*}
which does not have a closed analytical form. We use the Metropolis-Hastings algorithm with a lognormal proposal, whose
mean of the associated normal is the logarithm of the current value of the parameter, and the variance is previously tuned
to provide acceptance rates between 25\% and 45\%.	
	
	However, in the particular case when $\tau = 1/2$, $	p(\sigma^{-2}_j | \dots )$ has an analytical closed form and:
	\begin{equation*}
	\sigma^{-2}_j | \cdot  \sim Gama \left( \frac{n + \nu}{2} , \frac{\nu s^2}{2} +  \frac{\tau(1-\tau)}{4} \displaystyle \sum_{i=1}^n  \frac{(y_{ij} - \beta_j f_i)^2} {w_i} \right) 
	\end{equation*}
	
Then, when $\tau = 1/2$, all the full conditional distributions present closed form and just the Gibbs sampler algorithm is used to obtain samples from $p(\Theta\mid\dots)$.

\section{Model comparison criteria}\label{sec:criterio-selecao-modelo}

In this section we briefly describe the model comparison criteria used to compare the fitted models in Section \ref{sec3}.

\subsection{Likelihood and information criteria}

Traditional model selection criteria based on the likelihood include variants of Akaike information criterion (AIC) \cite{akaike1974new}, the Bayesian information criteria (BIC) \cite{schwarz1978estimating}), and related information criteria such as the informational complexity (ICOMP) \cite{bozdogan1998bayesian}.

From equation in (\ref{eq:definicao-mfq}), we have that, marginal on latent factors:
\begin{equation*}
    y_i |  \beta, \Delta \sim N_p(m w_i, \Lambda), 
\end{equation*}
where $\Lambda = \beta \beta' + w_i \Delta$. It is easily deduced that the likelihood function is given by:
\begin{equation*}
    L = \prod_{i=1}^n |\Lambda|^{-1/2} (2\pi)^{-p/2} exp \left[ -\frac{1}{2} (y_i - m w_i)'\Lambda^{-1} (y_i - m w_i) \right].
\end{equation*}
Define $l = -2log(L)$. Then, the various model selection criteria are defined as follows: (i) $AIC = l + 2 p_k$, (ii) $BIC = l + log(n) p_k$, (iii)  $BIC^* = l + log(\tilde{n}) p_k$ e (iv) $ICOMP = l + g( \hat{\Delta})$, where 
\begin{equation*}
    p_k = p(k+1) - \frac{k}{2}(k-1), \,\,
    \tilde{n} = n - \frac{2p + 11}{6} - \frac{2k}{3} \textrm{ and }
\end{equation*}
\begin{equation*}
    g(\Delta) = 2(k+1) \left[\frac{p}{2}log \left(trace(\Delta) \frac{1}{m}\right) - \frac{1}{2}log |\Delta|    \right].
\end{equation*}

Smaller values of these criteria indicate the best model among the fitted ones.

\subsection{Ranked probability scores}

\cite{gneiting2007probabilistic} considered scoring rules to assess the quality of probabilistic forecasts. In particular, the continuous
ranked probability score (CRPS) is computed as follows. For each $y_i$, the RPS can be expressed as:
\begin{equation*}
    RPS = \frac{1}{pn} \displaystyle \sum_{j=1}^p  \sum_{i=1}^n \mathbb{E} \left[ | y_{\textrm{rep},i} - y_i | \right] - \frac{1}{2} \mathbb{E} \left[ | y_{\textrm{rep},i} - \tilde{y}_{\textrm{rep},i} | \right],
\end{equation*}

where $y_{\textrm{rep},i}$ and  $\tilde{y}_{\textrm{rep},i}$ are independent replicates from the posterior predictive distribution. Assuming there is a sample of size $T$ from the posterior distribution of the parameters in the model, the previous expectations are approximated by $\frac{1}{T}\sum_{t=1}^T | y_{\textrm{rep},i}^{(t)} - y_i |$ e 
$\frac{1}{T}\sum_{t=1}^T | y_{\textrm{rep},i}^{(t)} - \tilde{y}_{\textrm{rep},i}^{(t)} | $. Smaller values of RPS indicate the best model among the fitted ones.

\subsection{Mean absolute error and mean square error} 
Standard measures of goodness of fit were also entertained in this study for comparison purposes. They are the mean square
error (MSE) and the mean absolute error (MAE):

\begin{equation*}
    MSE = \frac{1}{pn} \displaystyle \sum_{j=1}^p  \sum_{i=1}^n \left( y_i - \hat{y_i} \right) ^2  \textrm{    and    } 
    MAE = \frac{1}{pn} \displaystyle \sum_{j=1}^p  \sum_{i=1}^n | y_i - \hat{y_i} |,
\end{equation*}

where $\hat{y}_i$ is obtained through a Monte Carlo estimate of the posterior mean of the predictive distribution, across $T$ draws. Smaller values of MSE and MAE indicate the best model among the fitted ones.

\bibliographystyle{unsrt}  


\begin{thebibliography}{10}

\bibitem{lopes2004bayesian}
Hedibert~Freitas Lopes and Mike West.
\newblock Bayesian model assessment in factor analysis.
\newblock {\em Statistica Sinica}, pages 41--67, 2004.

\bibitem{wedel2003factor}
Michel Wedel, Ulf B{\"o}ckenholt, and Wagner~A Kamakura.
\newblock Factor models for multivariate count data.
\newblock {\em Journal of Multivariate Analysis}, 87(2):356--369, 2003.

\bibitem{cagnone2012factor}
Silvia Cagnone and Cinzia Viroli.
\newblock A factor mixture analysis model for multivariate binary data.
\newblock {\em Statistical Modelling}, 12(3):257--277, 2012.

\bibitem{zhang2014robust}
Jianchun Zhang, Jia Li, and Chuanhai Liu.
\newblock Robust factor analysis using the multivariate t-distribution.
\newblock {\em Statistica Sinica}, 24(1):291--312, 2014.

\bibitem{castro2015likelihood}
Luis~Mauricio Castro, Denise~Reis Costa, Marcos~Oliveira Prates, and
  Victor~Hugo Lachos.
\newblock Likelihood-based inference for tobit confirmatory factor analysis
  using the multivariate student-t distribution.
\newblock {\em Statistics and Computing}, 25(6):1163--1183, 2015.

\bibitem{CapdevilleGoncPer}
V~Capdeville, K~Gon\c{c}alves, and J~Pereira.
\newblock Bayesian factor models for multivariatecategorical data applied to
  questionnaires.
\newblock {\em Technical report}, 2019.

\bibitem{tobias2016covar}
Adrian Tobias and Markus~K Brunnermeier.
\newblock Covar.
\newblock {\em The American Economic Review}, 106(7):1705, 2016.

\bibitem{villarini2011frequency}
Gabriele Villarini, James~A Smith, Mary~Lynn Baeck, Renato Vitolo, David~B
  Stephenson, and Witold~F Krajewski.
\newblock On the frequency of heavy rainfall for the midwest of the united
  states.
\newblock {\em Journal of Hydrology}, 400(1-2):103--120, 2011.

\bibitem{koenker1978regression}
Roger Koenker and Gilbert Bassett~Jr.
\newblock Regression quantiles.
\newblock {\em Econometrica: journal of the Econometric Society}, pages 33--50,
  1978.

\bibitem{yu2001bayesian}
Keming Yu and Rana~A Moyeed.
\newblock Bayesian quantile regression.
\newblock {\em Statistics \& Probability Letters}, 54(4):437--447, 2001.

\bibitem{kozumi2011gibbs}
Hideo Kozumi and Genya Kobayashi.
\newblock Gibbs sampling methods for bayesian quantile regression.
\newblock {\em Journal of statistical computation and simulation},
  81(11):1565--1578, 2011.

\bibitem{kotz2012laplace}
Samuel Kotz, Tomasz Kozubowski, and Krzystof Podgorski.
\newblock {\em The Laplace distribution and generalizations: a revisit with
  applications to communications, economics, engineering, and finance}.
\newblock Springer Science \& Business Media, 2012.

\bibitem{teamR}
{R Core Team}.
\newblock {\em R: A Language and Environment for Statistical Computing}.
\newblock R Foundation for Statistical Computing, Vienna, Austria, 2017.

\bibitem{plummer2006coda}
Martyn Plummer, Nicky Best, Kate Cowles, and Karen Vines.
\newblock Coda: convergence diagnosis and output analysis for mcmc.
\newblock {\em R news}, 6(1):7--11, 2006.

\bibitem{choi2018quantile}
Ji-Eun Choi and Dong~Wan Shin.
\newblock Quantile correlation coefficient: a new tail dependence measure.
\newblock {\em arXiv preprint arXiv:1803.06200}, 2018.

\bibitem{citation_faraway}
Julian Faraway.
\newblock {\em faraway: Functions and Datasets for Books by Julian Faraway},
  2016.
\newblock R package version 1.0.7.

\bibitem{jorgensen2012statistical}
Bent Jorgensen.
\newblock {\em Statistical properties of the generalized inverse Gaussian
  distribution}, volume~9.
\newblock Springer Science \& Business Media, 2012.

\bibitem{akaike1974new}
H~Akaike.
\newblock A new look at the statistical identification model.
\newblock {\em IEEE Transactions on Automatic Control}, 19:716, 1974.

\bibitem{schwarz1978estimating}
Gideon Schwarz et~al.
\newblock Estimating the dimension of a model.
\newblock {\em The annals of statistics}, 6(2):461--464, 1978.

\bibitem{bozdogan1998bayesian}
Hamparsum Bozdogan and Kazuo Shigemasu.
\newblock Bayesian factor analysis model and choosing the number of factors
  using a new informational complexity criterion.
\newblock In {\em Advances in data science and classification}, pages 335--342.
  Springer, 1998.

\bibitem{gneiting2007probabilistic}
Tilmann Gneiting, Fadoua Balabdaoui, and Adrian~E Raftery.
\newblock Probabilistic forecasts, calibration and sharpness.
\newblock {\em Journal of the Royal Statistical Society: Series B (Statistical
  Methodology)}, 69(2):243--268, 2007.

\end{thebibliography}

\end{document}